\documentclass[fleqn,usenatbib]{mnras}

\usepackage{newtxtext,newtxmath}

\usepackage[T1]{fontenc}

\DeclareRobustCommand{\VAN}[3]{#2}
\let\VANthebibliography\thebibliography
\def\thebibliography{\DeclareRobustCommand{\VAN}[3]{##3}\VANthebibliography}

\usepackage{graphicx}	
\usepackage{amsmath}	

\title[HD~5501]{HD 5501: A Rapidly Evolving Interacting Eclipsing Binary with a Variable Light Curve and H~$\alpha$ Emission}

\author[R. O. Gray et al.]{
Richard O. Gray,$^{1}$\thanks{E-mail: grayro@appstate.edu}
Christopher. J. Corbally,$^{2}$
Sean Curry,$^{3}$
Bradley E. Schaefer,$^{4}$
Jack Martin,$^{5}$
David Boyd,$^{6}$
\newauthor
James Foster,$^{7}$
Dale E. Mais,$^{8}$
Michael M. Briley,$^{1}$
Forrest Sims,$^{9}$
Christophe Boussin,$^{10}$
Gary Walker,$^{11}$
\newauthor
Joe Novosel,$^{12}$
David Cejudo Fernandez,$^{13}$
Robert Buchheim,$^{14}$
David Iadevaia,$^{15}$
Robin Leadbeater,$^{16}$
\newauthor
Daniel B. Caton,$^{1}$
Adam Smith,$^{17}$
Courtney E. McGahee,$^{18}$
David Decker,$^{19}$
and Gary Hawkins $^{20}$
\\
$^{1}$Department of Physics and Astronomy, Appalachian State University
  Boone, NC 28608, USA\\
$^{2}$Vatican Observatory Research Group, University of Arizona,  Tucson, AZ 85721-0065 USA\\
$^{3}$Yank Gulch Observatory, Talent, OR 97540 USA\\
$^{4}$Department of Physics and Astronomy, Louisiana State University, Baton Rouge, LA 70803, USA\\
$^{5}$Huggins Spectroscopic Observatory, Rayleigh, Essex, SS6 8AW, UK\\
$^{6}$BAA Variable Star Section, West Challow Observatory, West Challow, OX12 9TX, UK\\
  $^{7}$AAVSO/ARAS, Pinon Pines RoR Observatory,  Frazier Park, CA 93225, USA\\
  $^{8}$14690 Waterstradt Rd., Marcellus, MI. 49067, USA\\
  $^{9}$Desert Celestial Observatory, Apache Junction, AZ 85119, USA\\
  $^{10}$Observatoire de l'Eridan et de la Chevelure de B\'er\'enice, 02400 \'Epaux-B\'ezu , France\\
  $^{11}$Maria Mitchell Observatory, Minor Planet Center \#811, 4 Vestal Street, Nantucket, MA 02554, USA\\
  $^{12}$11623 Tillbury Cv, Fort Wayne IN 46845, USA\\
  $^{13}$Camino de las Canteras, 42, El Berruero 28192, Spain\\
  $^{14}$Lost Gold Observatory, 8731 E. Lost Gold Cir, Gold Canyon, AZ
  85118, USA\\
  $^{15}$Mountain View Observatory, 5700 N Avenida Observatory, Tucson, AZ 85750 USA\\
  $^{16}$Three Hills Observatory, The Birches, CA7 1JF, UK\\
  $^{17}$Gemini North Observatory, Hilo Hawaii, USA\\
  $^{18}$Lenoir-Rhyne University, Engineering Physics, Hickory, NC 28601, USA\\
  $^{19}$4007 S. Tropico Dr., La Mesa, CA 91941, USA\\
  $^{20}$Blossom Valley Small Telescope Observatory, El Cajon, CA 92021, USA}

\date{Accepted XXX. Received YYY; in original form ZZZ}

\pubyear{2025}

\begin{document}
\label{firstpage}
\pagerange{\pageref{firstpage}--\pageref{lastpage}}
\maketitle

\begin{abstract}
HD~5501, a hitherto little studied eclipsing binary with an early A-type
primary, has been caught in a short-lived, astrophysically interesting phase
of its binary evolution.  Recent photometric and spectroscopic observations, including photometric data from {\it TESS}, show it has a highly variable light curve as well as complex spectral variability, particularly in both the absorption and emission components at H~$\alpha$. Our current campaign, including both professional and amateur observers, has determined that the primary is evolving rapidly across the Hertzsprung gap and that, unusually in the case of mass transfer, the orbital period is declining with a characteristic time-scale $P/\dot{P} \approx$ 170,000 years. Significantly, the orbit is eccentric and it appears that mass transfer from the primary to the secondary occurs only near periastron. Modeling indicates the presumed B7 V secondary to be surrounded by an accretion torus, which likely has dynamically chaotic variations in size and shape. Our analysis further implies the presence of a circumbinary disc or shell supplied by mass loss through the Lagrange $L_3$ point. That mass loss appears to account for most of the emission at H~$\alpha$. We describe how this astrophysically interesting system may yield valuable information about binary star evolution at the onset of Roche-lobe overflow, as well as insights into eccentricity-modifying mechanisms such as the Soker mechanism. 
\end{abstract}




\section{Introduction}
\label{sec:intro}

HD~5501 $=$ BD$+59^\circ 154 = $ BSD 8-397 is a little-studied 9th magnitude
eclipsing binary system in Cassiopeia with an early A-type primary.  It was
first classified spectroscopically during the Bergedorfer
Spektral-Durchmusterung \citep{schwassmann1935} objective-prism survey of the
Kapteyn selected regions.  Designated BSD 8-397, it was given a spectral type
of B7p with a note that the K-line was unusually strong.  Later,
\citet{hardorp1959}, in the course of the Luminous Stars in the Northern
Milky Way survey, gave it a spectral type of A0 Ib.  \citet{fehrenbach1961}
and \citet{barbier1968} observed it during an objective-prism radial-velocity
survey, estimated a spectral type of A3 II, and measured a radial velocity of
$-46$ km s$^{-1}$.  Apart from observations on the Johnson {\it UBV} system
\citep{bigay1965} and the Str\"{o}mgren {\it uvby}$\beta$ system
\citep{perry1982}, the literature is silent on the nature of this star apart
from a preliminary report we published based on early observations \citep{Mais2006}.

\begin{figure} 
\includegraphics[width=3.3in]{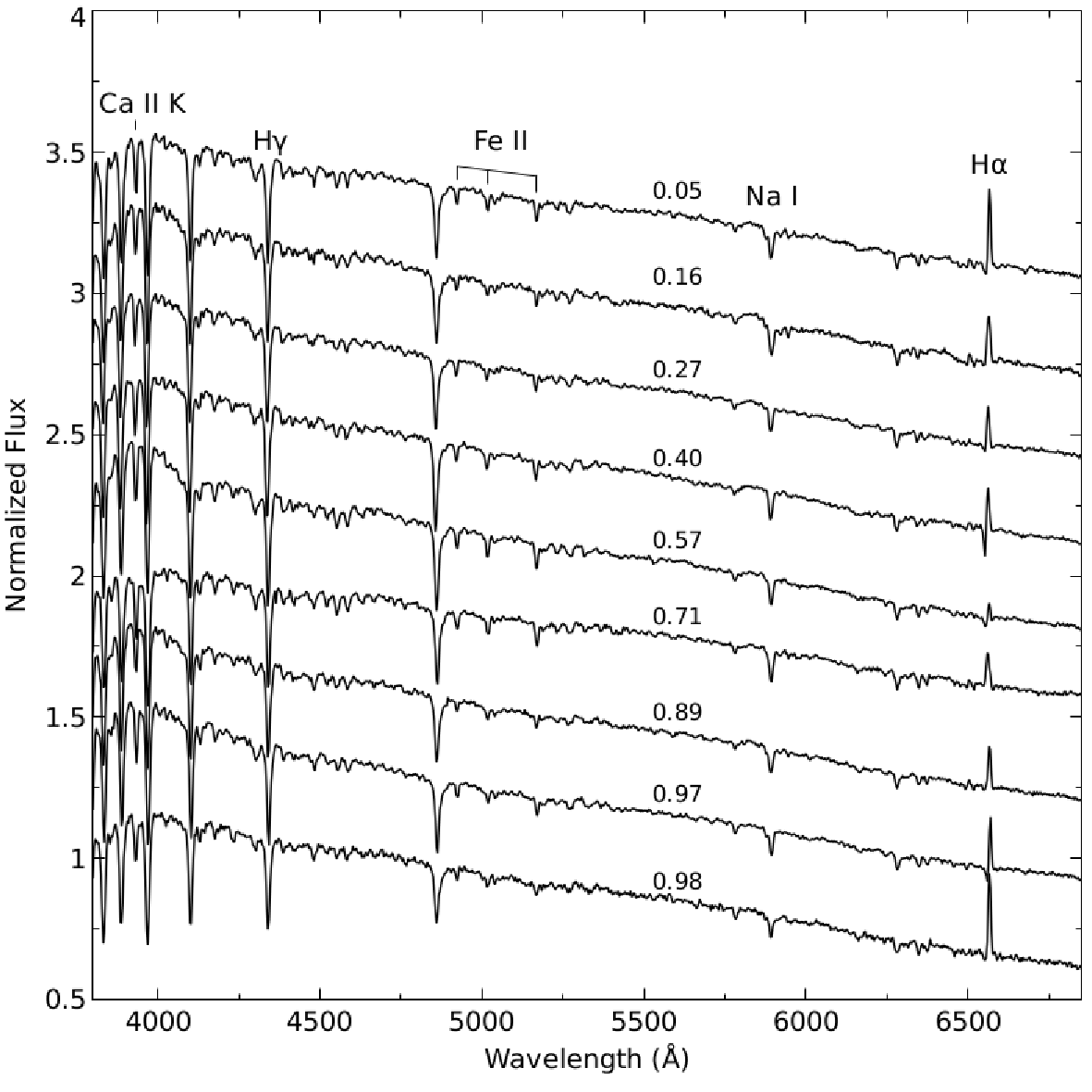} 
\caption{A montage of low-resolution spectra of HD~5501 ordered according to
  orbital phase (see \S \ref{sec:orbit}: primary eclipse at phase 0,
  secondary eclipse near 0.5). These spectra are flux
  calibrated, but normalised at a consistent wavelength and displaced by
  0.3 flux units for ease of comparison.  Certain spectral features are
  labelled, including H~$\alpha$, the most prominent shell lines, notably those
  of Fe~{\sc ii} multiplet 42 (labelled Fe~{\sc ii}), and the
  Ca~{\sc ii} K line.  All of these lines show variability.  Spectra obtained by
C. Boussin during the 2024/2025 observing season.}
\label{fig:LRM}
\end{figure}

HD~5501 was first observed spectroscopically by one of us (ROG) from the
Dark Sky Observatory (DSO) in early 2004 November during a spectral
classification
survey of late-B and early A-type stars in the BSD 8 and 9 regions.  The goal
of that unpublished survey was to discover new examples of A-type shell stars.
That first spectrum (apparently the first-ever slit spectrogram of the star)
revealed the characteristic spectral features of an A-type shell star, and so
a second spectrum was obtained in late 2004 November.  That spectrum confirmed
the shell star classification, but the Fe~{\sc ii} shell lines had strengthened and
the Balmer lines appeared more shallow.  Such variability is unusual (except
for a few notable exceptions) in the class of A-type shell stars, and so
further spectroscopic observations were scheduled.  Those observations, which
will be reviewed in \S \ref{sec:spt}, show a clear variation in the Balmer-line
profiles, the strength of the Fe~{\sc ii} shell lines, and the Ca~{\sc ii}
K-line with a
period of about 7.5 days (see Figure \ref{fig:LRM}).

\begin{figure} 
\includegraphics[width=2.5in,angle=-90]{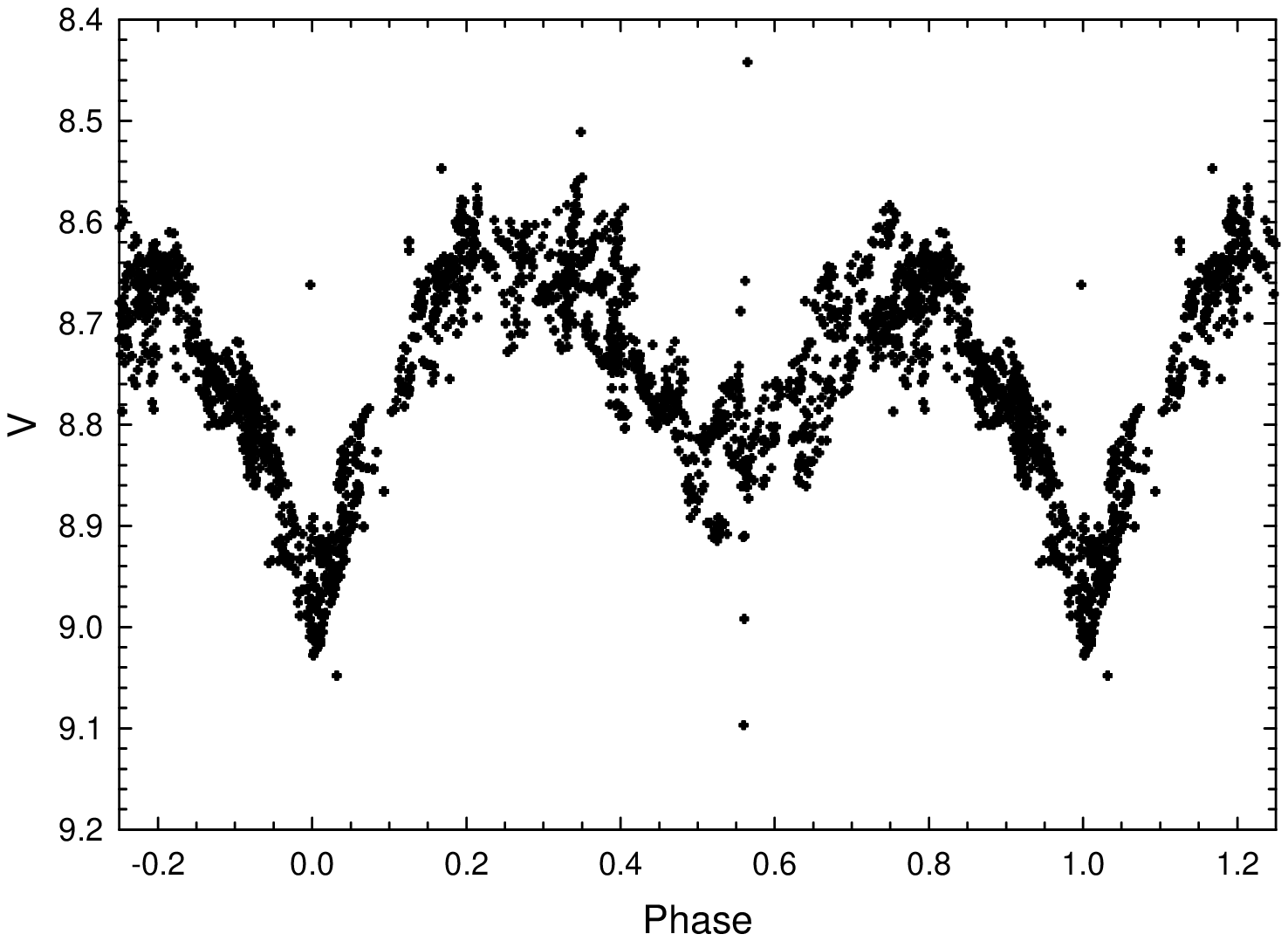} 
\caption{A preliminary Johnson-$V$ lightcurve for HD~5501, based on early
  photometry obtained by Dale Mais and observers at the Dark Sky Observatory.
  The observations have been phased with a preliminary period, based on these
data alone of 7.5338 days.}
\label{fig:LC1}
\end{figure}

As it turned out, HD~5501 had also been observed photometrically during two
observing seasons (2003-2004) by The Amateur Sky Survey
\citep[TASS][]{Droege2006}.
That photometry shows clear
variability with an amplitude of $\approx 0.35$ mag in the Johnson $V$-band.
When phased with the period of the spectral variability, the light curve
shows a primary and secondary eclipse, but with an unusual amount of ``noise''.
That discovery prompted photometric observations of HD~5501 at
DSO, and we also enlisted amateur Dale Mais for photometric observations
from his
private observatory.  Figure \ref{fig:LC1} shows the combined Johnson-$V$
light curve from
the DSO and Mais photometry, phased with a preliminary period based on those
data
of 7.5338 days.  Those data confirmed that HD~5501 is an eclipsing binary
with a light curve that appears to show ``noise'' that cannot be explained
by photometric errors.  This result will be examined in more detail in sections
\ref{sec:orbit}, \ref{sec:chaos}, and \ref{sec:lcurve}.

These initial observations indicated that HD~5501 was an object of considerable
interest.  However, the pressure of other observing projects relegated
HD~5501 to the back burner until the Fall of 2023 when an opportunity arose
to obtain extensive photometric and spectroscopic observations in collaboration
with a group of capable and enthusiastic amateurs.

\section{Observations}

\subsection{Photometry}
\label{sec:phot}

\begin{table*}
  \caption{Sources of Photometry}
  \label{tab:BVR}
  \resizebox{\textwidth}{!}{%
  \begin{tabular}{lcllll}
    \hline
    \hline
    Observatory & Telescope Diam (m) & Filters & Location & Observer & Abbreviation\\
    \hline
    Dark Sky Observatory & 0.8 & $B$, $V$, $R_C$ & North Carolina & Adam Smith \& Daniel Caton & DSO32 \\
    Dark Sky Observatory & 0.08 & S-{\it v}, $B$, $V$, $R_C$, H~$\alpha$ & North Carolina & Gray, Briley \& McGahee & DSOWF \\
    Mais Observatory  & 0.13 & $B$, $V$, $R_J$, $I_J$ & California & Dale Mais & DEM\\
    West Challow Observatory & 0.36 & $B$, $V$, $R_C$ & UK & David Boyd & BDG\\
    Observatorio El Gallinero & 0.07 & $B$, $V$ & Spain & David Cejudo Fernandez & CDZ\\
    Desert Celestial Observatory & 0.1  & $V$ & Arizona  & Forrest Sims  & SFOA\\
    Sierra Remote Observatory & 0.51 & $U$, $V$, H~$\alpha$ & California & Gary Walker & WGR\\
    Pinon Pines RoR Observatory & 0.36 & $U$, $B$, $V$, $I$ & California & James Foster & JRF\\
    \hline
  \end{tabular}}
\end{table*}

As mentioned in the introduction, early CCD photometry was obtained for HD~5501 at the Dark Sky Observatory using
the 0.8-m reflector and by Dale Mais at his personal observatory.  Both those sets of photometry were obtained
during the observing season for HD~5501 in 2004.  Photometry of HD~5501 in the $B$, $V$, $R_C$ bands as well as a 3-nm wide
filter centred at H~$\alpha$ recommenced at the Dark Sky Observatory in 2018 October using the Robotic Wide Field
Instrument (DSOWF).  That instrument consists
of a 300mm f/4 telephoto lens equipped with a Finger Lakes camera (KAF-16803 chip) and filter wheel.  The data from
this instrument are reduced with a pipeline using functions from the
python {\sc ccdproc} package
\citep{matt_craig_2017_1069648}.  Differential aperture photometry is carried out with functions from the python
library {\sc photutils} \citep{larry_bradley_2023_7946442}.  Five exposures in each filter were obtained on each clear
night HD~5501 could be observed, but that frequency was increased during the 2023/24 observing season during which
exposures in the Str\"{o}mgren-$v$ filter were added.  The comparison (HD~236580) and check (BD+59 159) stars were
observed on the same images.  Table \ref{tab:BVR} also lists the amateur observatories (BDG, CDZ, SFOA, WGR and JRF) that
obtained extensive photometry for HD~5501 during the 2023-24 and 2024-25
campaigns.  Those observations may be found in the
AAVSO database\footnote{\url{https://www.aavso.org/databases}}.  With one exception, those observers used the same
comparison star, but in some cases a different check star from the one employed by the DSO robotic wide field instrument (DSOWF).

Only {\it Tycho-2} \citep{Hog2000} $B$ and $V$ photometry are available for the comparison star HD~236580.  {\it Tycho-2}
photometry is only approximately on the Johnson $BV$ system \citep[see discussion in][]{Bessell2000}.  The
star HD~5015, located about one degree from HD~5501 and its
comparison star, has been observed on the Johnson $UBVR$ system \citep{ducati2002}.  That star, unfortunately, is saturated on the
images we obtained for HD~5501, so on a number of nights during December 2023 we used the DSOWF instrument
to obtain images centred alternately on HD~5501 and HD~5015.  Those observations yielded the following magnitudes
for HD~236580: $B = 9.56 \pm 0.016$, $V = 8.52 \pm 0.01$, $R_C = 7.92 \pm 0.02$.  Since HD~5015 was observed with
a Johnson $R$ filter and the Wide Field instrument employs a Cousins $R$ filter, we transformed the $R_J$ magnitude
of HD~5015 to an $R_C$ magnitude using the transformation of \citet{Bessell83}.  Our derived $B$ and $V$ magnitudes
for HD~236580 are reasonably close to the {\it Tycho-2} magnitudes ($B = 9.64$, $V = 8.56$).  Since HD~5015 was also
observed on the Str\"{o}mgren $uvby$ system, we were able to determine a Str\"{o}mgren-$v$ magnitude for HD~236580.
We obtained: S-$v = 10.38 \pm 0.02$.

WGR also obtained Johnson-$U$ photometry for HD~5501 and the
comparison star HD~236580 from the Sierra Remote Observatory at
an elevation of 1400~m.  Furthermore, $U$-band observations of HD~5015 and the
HD~5501 field (including the comparison star HD~236580) were obtained by JRF
from the Pinon Pines RoR Observatory at an elevation of 1685~m.  This
enabled the transfer of
$U$ photometry of HD~5015 to the comparison star and hence to HD~5501.  That
exercise yielded $U({\rm HD~236580}) = 10.48 \pm 0.05$.  As it turns out,
this is reasonably close to the $U$ magnitude expected for a star at the
same distance, reddening \citep[from the reddening map of ][]{green2019} and
spectral type as HD~236580 (G5 II-III) based on tabular data in \citet{cox00}:
$U = 10.32 \pm 0.05$.  We employed $U = 10.48$ to place the $U$ photometry
for HD~5501 on the standard Johnson system.  

\subsection{Spectroscopy}

Classification-resolution spectroscopy of HD~5501 has been
obtained sporadically using the GM spectrograph on the 0.8-m telescope at the
Dark Sky Observatory since 2004, and then on a more systematic basis since
2022.  HD~5501 has been observed with both the 1200 g mm$^{-1}$ and the
600 g mm$^{-1}$ gratings (yielding resolutions of $R \approx 3000$ and
$R \approx 1300$
and spectral ranges of $3800 - 4700$\AA\ and $3800 - 5600$\AA\ respectively).
 These spectra
 were reduced with a custom python pipeline which uses functions from
 {\sc astropy}
\citep{astropy2013,astropy2018,astropy2022} and {\sc ccdproc} \citep{matt_craig_2017_1069648}.  The pipeline employs optimal extraction of the spectra using the
algorithm from \citet{horne86} and detailed modeling of the sky background.
Approximate
flux calibration was carried out through the observation of several
spectrophotometric standards most nights.  Wavelength calibrations were based
on observations of a hollow-cathode FeAr lamp both before and after the
science exposures.

\begin{table*}
  \caption{Sources of Spectroscopy}
  \label{tab:spectra}
  \resizebox{\textwidth}{!}{%
  \begin{tabular}{llllllll}
    \hline
    \hline
    Observatory & Telescope Diam (m) & Spectrograph & Resolution & Spectral Range (\AA) & Location & Observer & Abbreviation\\
    \hline
    Dark Sky Observatory & 0.80 & GM spectrograph & 1300 & 3800 -- 5700 & North Carolina & Gray, Briley \& McGahee & DSO\\
    Dark Sky Observatory & 0.80 & GM spectrograph & 3000 & 3800 -- 4700 & North Carolina & Gray, Briley \& McGahee & DSO\\
    Vatican Observatory & 1.8 & VATTspec & 3000 & 3750 -- 5500 & Arizona & Christopher Corbally & VATT\\
    Sta. Maria de Montmagastrell & 0.41 & NOU\_T Echelle & 9000 & 3840 -- 8150 & Spain & Sean Curry & SMM\\
    Huggins Spectroscopic Observatory & 0.36 & Shelyak LHIRES III & 7000 & 6320 -- 6595 & UK & Jack Martin & HSO\\
    Observatoire de l'Eridan et & 0.20 & Shelyak ALPY 600 & 530 & 3700 -- 7565 & \'Epaux-B\'ezu, France & Christophe Boussin & OECB\\
    \phantom{O}\phantom{O}de la Chevelure de B\'er\'enice & & & & & & \\
    West Challow Observatory & 0.28 & Shelyak LISA & 1000 & 3900 -- 7400 & West Challow, UK & David Boyd & WCO\\
    West Challow Observatory & 0.28 & Shelyak StarEx & 10500 & 6260 -- 6700 & West Challow, UK & David Boyd & WCO\\ 
    Desert Celestial Observatory & 0.51 & Shelyak LISA & 1000 & 3750 -- 7300 & Arizona & Forrest Sims & DCO\\
    Observatorio El Gallinero &  0.36 & Shelyak LISA & 750 & 3950 -- 7380 & Spain & David Cejudo & CDZ\\
    Mountain View Observatory & 0.24  & ALPY 600 & 500 & 3800 -- 7500 & Arizona & David Iadevaia & IDG\\
    La Mesa Observatory & 0.20 & ALPY 600 & 600 & 3700 -- 7280 & California & David Decker & DDGB\\
    Star*Quest & 0.36 & Shelyak LHIRES III & 4237 & 6318 -- 6775 & Indiana & Joe Novosel & NSQ\\
    Pinion Pines RoR Observatory & 0.43 & Shelyak LHIRES III & 12000 & 6486 -- 6640 & California & James Foster & JRF\\
    Three Hills Observatory & 0.28 & Shelyak LHIRES III & 14000 & 6480 -- 6630 & UK & Robin Leadbeater & THO\\
    \hline
  \end{tabular}}
\end{table*}

In addition, spectra were obtained on the 1.8-m Vatican Observatory Advanced
Technology Telescope (Alice P. Lennon Telescope) employing the VATTspec, a 1
arcsecond slit, a
600 g mm$^{-1}$ grating and a STA0520A back-thinned CCD with 15$\mu$m pixels,
which yield a resolution of $R = 3000$ and a spectral range of
$3750 - 5500$\AA.
The VATTspec spectra were reduced with {\sc iraf} \citep{iraf86,iraf93} using
standard procedures, and were approximately flux calibrated via the observation
of a few spectrophotometric standards.

Spectra were also obtained at a number of amateur observatories listed in
Table \ref{tab:spectra}. These
spectra were reduced using various packages designed for amateur
use, including {\sc isis}
\footnote{\url{http://www.astrosurf.com/buil/isis-software.html}} and
{\sc bass} \footnote{\url{https://groups.io/g/BassSpectro}}.  Approximate flux calibration
was carried out
using observations of standard stars and placed on an absolute scale via
Johnson $V$ photometry \citep{boyd2020}.

\section{Analysis and Discussion}

HD~5501 is an eclipsing binary system that exhibits a number of unusual
features.  We shall demonstrate in this section that the orbit is evolving
on a rapid time-scale, with a decreasing period; the light curve is peculiar
in the sense that the shapes and depths of the primary and secondary eclipses
vary from one orbit to the next; that those light curve peculiarities suggest
the presence of dynamical chaos in the system; the profile of the H~$\alpha$ is
complex and highly variable, and shows both emission and absorption
components; both the complex H~$\alpha$ profile and the strengths of Fe~{\sc ii}
shell lines, as well as the profiles of other strong lines such as Ca~{\sc ii} K
suggest the presence of circumbinary material, with velocities up to 500 km
sec$^{-1}$ relative to the systemic velocity; and that the orbit of the
system is eccentric.

\subsection{The Orbital Period and Orbital Evolution}
\label{sec:orbit}

\begin{figure} 
\includegraphics[width=3.2in,angle=0]{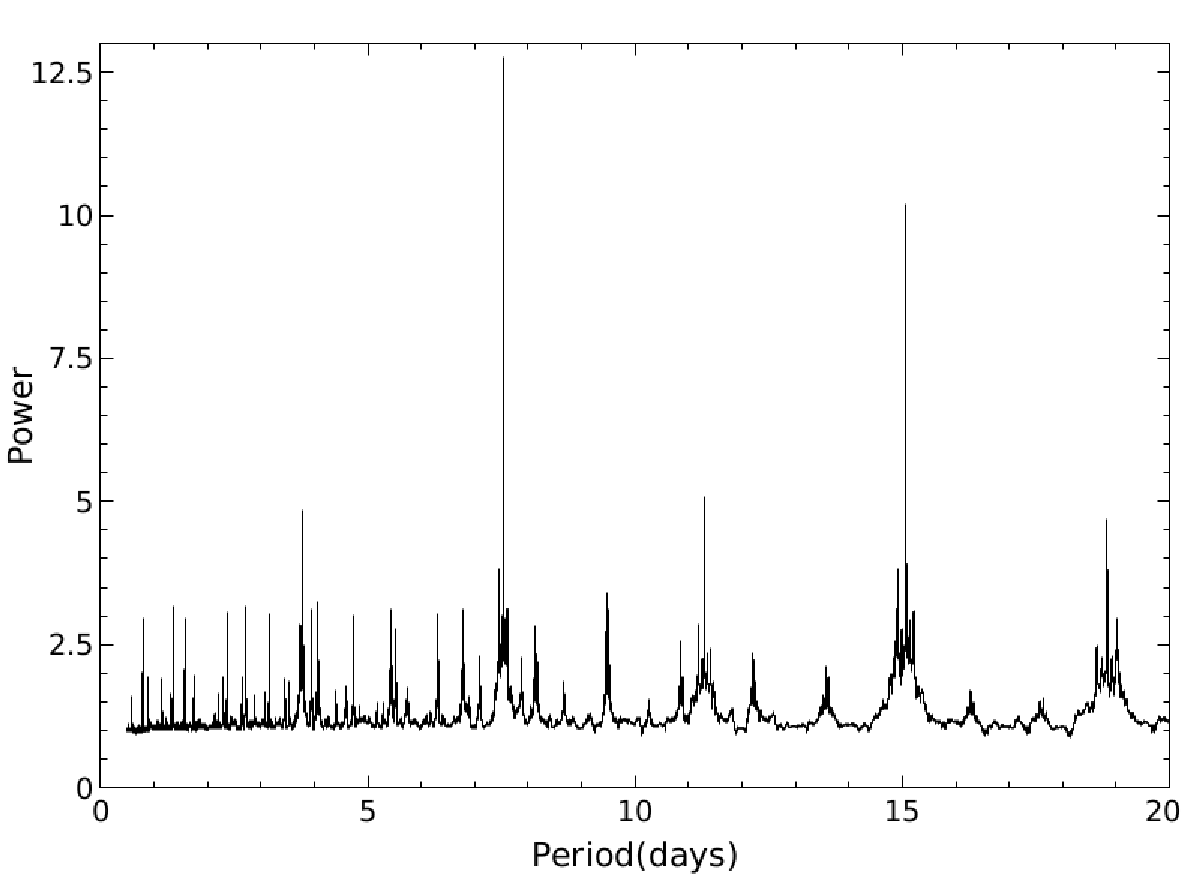} 
\caption{A periodogram computed with the Plavchan algorithm for the DSOWF 2018-2023 Johnson-$V$ photometry.  The first-ranked period is 7.53085 days.  Many of the other peaks are simple integer fractions of that period.}
\label{fig:plavchan}
\end{figure}

The homogeneous Johnson-$V$ photometry from the DSOWF Instrument
(2018 Oct -- 2023 Dec, 522 points) was used for an initial determination of the orbital
period of the binary system.  We employed the periodogram service of the NASA Exoplanet
Archive \footnote{https://exoplanetarchive.ipac.caltech.edu/cgi-bin/Pgram/nph-pgram} to conduct the period search.  That service offers the user a choice of
three period-finding algorithms: Lomb-Scargle, Box-fitting Least squares, and
Plavchan.  The second of those algorithms is specialized for the determination of
exoplanet transit periods.  The Lomb-Scargle algorithm, which is an adaptation of the
Discrete Fourier Transform, uses sinusoids as the basis functions, which is
not necessarily the best choice for an eclipsing binary.  Indeed, application of that
algorithm yields a first-ranked period of 3.7656 days, but inspection of the folded
data indicates that the algorithm does not distinguish between the primary and secondary
eclipses, and so the actual period is roughly twice that.  In the Plavchan algorithm \citep{Plavchan2008},
a variation of the phase dispersion minimization method \citep{Stellingwerf78}, the periodic basis
functions are essentially derived from the data.  The Plavchan algorithm easily
distinguishes between the primary and secondary minima of HD~5501, and yields a period of
$7.53085 \pm 0.00330$ days where the error is estimated from the HWHM of the first-ranked peak
(see Figure \ref{fig:plavchan}).  Many of the lower ranked peaks in that periodogram
correspond to simple integer fractions of that period.

A preliminary ephemeris, based only on the DSOWF photometry is given by
\begin{equation}
  T = 2460000.640 + 7.53085N
\end{equation}
where $N = $ integer gives the Julian Date ($T$) of a primary eclipse.

An important question to ask is whether this system shows orbital evolution.
Eight datasets, which span a period of nearly 120 years, can be used to examine
this question.
Those datasets are 1) the DSOWF 2018-2023 dataset referenced in previous
paragraphs, 2) the TASS Mark IV Photometric Survey of the Northern
Sky \citep{Droege2006} (2003-2004), 3) Dark Sky Observatory
DSO32 photometry (2004), 4) Mais Observatory photometry (2004), 5) the
Kamogata/Kiso/Kyoto wide-ﬁeld survey (KWS) archive \citep{Maehara2014}
(2012--2023), 6) {\it TESS} photometry (2019-2022), 7) the 2023-24 HD~5501 campaign photometry
carried out by AAVSO observers (BDG, CDZ, SFOA, and WGR -- see Table \ref{tab:BVR})
and 8) the Harvard Plate Archives \citep{Grindlay2012} which provide 1294
Johnson $B$ magnitudes from 1890 to 1989.

These datasets may be used to construct an ${\rm O} - {\rm C}$
(Observed $-$ Computed) diagram by finding dates for the primary minima
deduced from these datasets and comparing those with the preliminary ephemeris (equation 1)
above.  One complication, however, as will be seen below, is that the shape
and indeed the timing of the primary minimum varies from cycle to cycle.  As
a consequence, none of the datasets enumerated above (except for the {\it TESS} data)
have a sufficient density of points to allow determination of time of minimum for individual primary
eclipses.  Instead, an alternative approach is required.  In the case of, for
example, the KWS survey, the data are binned into two-year sets and phased
using the preliminary ephemeris, and then the {\it average} phase of the
primary minima is measured by shifting a model lightcurve in the phase domain
until the $\chi^2$ is minimized.  The error in the time of minimum was estimated from
the reduced $\chi^2$ values.  For datasets 1 -- 7, the model lightcurve was
derived by fitting a Fourier series to dataset (1).

The $B$ magnitudes from the
Harvard plates over a time interval (typically one decade) were fitted to a
periodic waveform with the eclipsing binary light curve shape, with the epoch
of minimum light determined from the minimum chi-square, and the one-sigma
error bars determined from the range of epochs that have a chi-square within
unity of that minimum.

\begin{figure} 
\includegraphics[width=3.2in,angle=0]{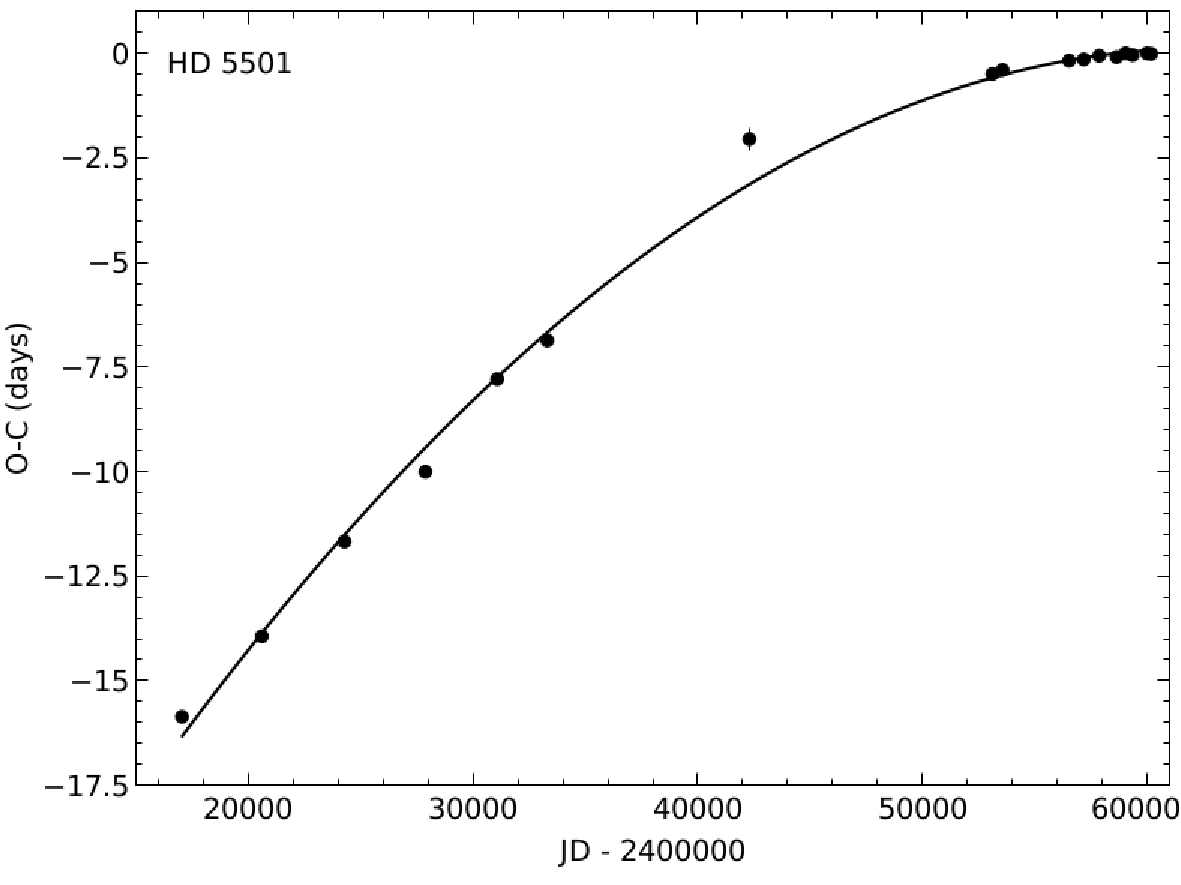} 
\caption{The ${\rm O} - {\rm C}$ diagram for the time of the primary eclipse
  of HD~5501 as a function of the date based on the ephemeris (Equation 1).
  The data are tabulated in Table \ref{tab:OC}. The curve is a weighted parabolic fit to the data points.  ``JD'' refers to the heliocentric Julian date. }
\label{fig:OC}
\end{figure}

The TASS dataset, which comprises data taken during two seasons (2003-4), was
placed in a single bin because of the small number of points.  Datasets (3) and (4) were
analysed together.  The results of the analysis of all the datasets are presented in Table \ref{tab:OC}
and illustrated in Figure \ref{fig:OC}.

{\small
\begin{table}
  \caption{Times of Primary Eclipse for HD~5501}
  \label{tab:OC}
  \resizebox{\columnwidth}{!}{%
  \begin{tabular}{lrrrc}
    \hline
    \hline
    Julian Date & N & ${\rm O} - {\rm C}$ (d) & $\sigma$ (d) & Dataset\\
    (Heliocentric) & & & & \\
    \hline
    2417043.871 & -5702 & -15.865 & 0.095 & 8 \\
    2420600.358 & -5230 & -13.940 & 0.093 & 8 \\
    2424277.678 & -4742 & -11.674 & 0.095 & 8 \\
    2427879.093 & -4264 & -10.006 & 0.089 & 8 \\
    2431074.648 & -3840 &  -7.790 & 0.073 & 8 \\
    2433304.445 & -3544 &  -6.866 & 0.156 & 8 \\
    2442301.095 & -2350 &  -2.051 & 0.265 & 8 \\
    2453116.954 &  -914 &  -0.492 & 0.081 & 2 \\
    2453568.902 &  -854 &  -0.395 & 0.008 & 3,4 \\
    2456528.747 &  -461 &  -0.174 & 0.044 & 5 \\
    2457183.959 &  -374 &  -0.146 & 0.031 & 5 \\
    2457869.355 &  -283 &  -0.057 & 0.021 & 5 \\
    2458645.001 &  -180 &  -0.089 & 0.047 & 5 \\
    2459044.227 &  -127 &   0.002 & 0.001 & 6 \\
    2459352.949 &   -86 &  -0.041 & 0.028 & 5 \\
    2460000.643 &     0 &   0.000 & 0.010 & 1 \\
    2460181.370 &    24 &  -0.011 & 0.018 & 7 \\
    \hline
    \end{tabular}}
\end{table}
}

It is clear that the ${\rm O} - {\rm C}$ (measured in units of days)  value has become more positive
with time in a non-linear way.  A linear but sloping ${\rm O} - {\rm C}$ curve would
simply mean that the period is constant but incorrect.  A non-linear ${\rm O} - {\rm C}$ curve
implies a change in the orbit which may arise from certain physical mechanisms including
1) a third body in the
system causing the line of apsides to rotate, leading to a change in the timing of
the eclipses, 2) a change in the orbital period due to mass transfer between the
components or mass loss from the system or 3) tidal forces between the two components
causing the orbit to evolve.

Apsidal motion or a third body would cause the ${\rm O} - {\rm C}$ curve
to be approximately sinusoidal \citep{Eggleton2006}.
However, the best fit to our data is parabolic.  To determine the time-scale for the orbital evolution
of the system, we have performed a weighted quadratic fit to the data in
Table \ref{tab:OC}.  However, for this system, it is important to take into account that there is
an inherent ``jitter'' in the timing of eclipses (we will discuss this in more detail in Section
\ref{sec:chaos}) when performing that fit.  We proceed by introducing an additional $\sigma_{\rm jitter}$
that is added in quadrature to the $\sigma$'s in Table \ref{tab:OC}.  The $\sigma_{\rm jitter}$, which is
assumed to be the same for each point in Table \ref{tab:OC} is adjusted until the reduced
$\chi^2 \approx 1$.  We find $\sigma_{\rm jitter} = 0.32$ days.  The quadratic fit may be used to derive
a revised ephemeris of the form:
\begin{equation}
  T = E_0 + P_0N + \tfrac{1}{2}P_0\dot{P}N^2
\end{equation}
where $P_0$ is the period at epoch $E_0$, and $\dot{P}$ is the
period derivative with units of (days/day).  Again the time, $T$, of a primary eclipse corresponds to
an integer value for $N$, the cycle number.
We derive: $E_0 = 2460000.710 \pm 0.135$ (JD), $P_0 = 7.531151 \pm 0.000187$ days, and
$\dot{P} = -1.200 \pm 0.095 \times 10^{-7}$.  This works out to $\dot{P} = -3.78 \pm 0.30$~s~yr$^{-1}$. The implied time-scale for the orbital
evolution $P/\dot{P} \approx $ 170,000 years is remarkably rapid.  We will discuss the significance
of this rapid time-scale in Section \ref{sec:parameters}.  We also note that the period is
{\it decreasing} with time, another unusual result.

Some Algol systems do show negative period derivatives. Table 4 in \citet{Erdem2014} lists values of $\dot{P}$ for eight Algols which show decreasing periods.  Of those eight, seven have
  $\dot{P} > -0.58$~s~yr$^{-1}$.  Only one, SX~Cas, shows a period derivative ($-2.6$~s~yr$^{-1}$) comparable to that of HD~5501.  In Algol systems a decreasing period is believed to be associated with magnetic braking due to the stellar wind flowing from the cool late-type component \citep[see][]{Erdem2014}. However, since in HD~5501 the mass transfer is from the initial primary to the secondary (see
  discussion in \S \ref{sec:parameters} and following), and both appear to be
  early-type stars, we do not believe that HD~5501 is an Algol system.       

\begin{figure} 
\includegraphics[width=2.6in,angle=0]{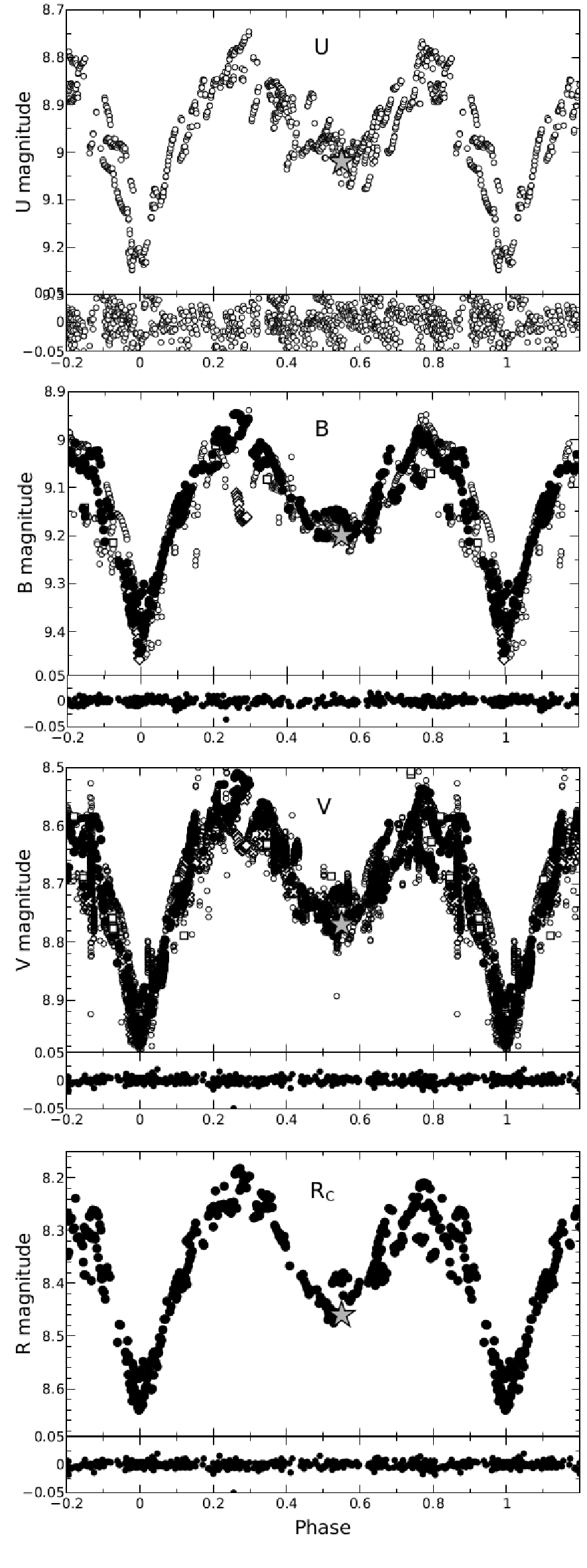} 
\caption{The phased HD~5501 light curves for the 2023-2024 observing season
  in the filters $U$, $B$, $V$ and Johnson-Cousins $R_C$.  The symbols
  represent the different sources for the photometry: $\bullet$: DSOWF; $\circ$: WGR; $\square$: BDG;
  $\diamond$: CDZ; $\times$: SFOA (see Table \ref{tab:BVR}).  The large gray
  star
  symbols indicate the adapted ``minimum'' magnitudes for the centres of the
  secondary eclipses.  Each panel
  shows
  the light curve above and photometric residuals (comparison $-$ check) below.
  The residuals are very large for the $U$ photometry because of the faintness
  of the check star in $U$.  The errors in the $U$ photometry are much smaller
  than implied by those residuals.  For the sake of clarity, only the residuals
  from DSOWF photometry are shown in the panels for $B$, $V$ and $R_C$.}
\label{fig:UBVR}
\end{figure}

$U$, $B$, $V$ and $R_C$ phased light curves for the 2023 - 2024 observing season, based on observations
from DSOWF, BDG, CDZ, SFOA and WGR (see Table \ref{tab:BVR}) are shown in Figure \ref{fig:UBVR}.

\subsection{{\it TESS} photometry and the peculiarities of the light curve}
\label{sec:chaos}

\begin{figure}
  \includegraphics[width=3.2in]{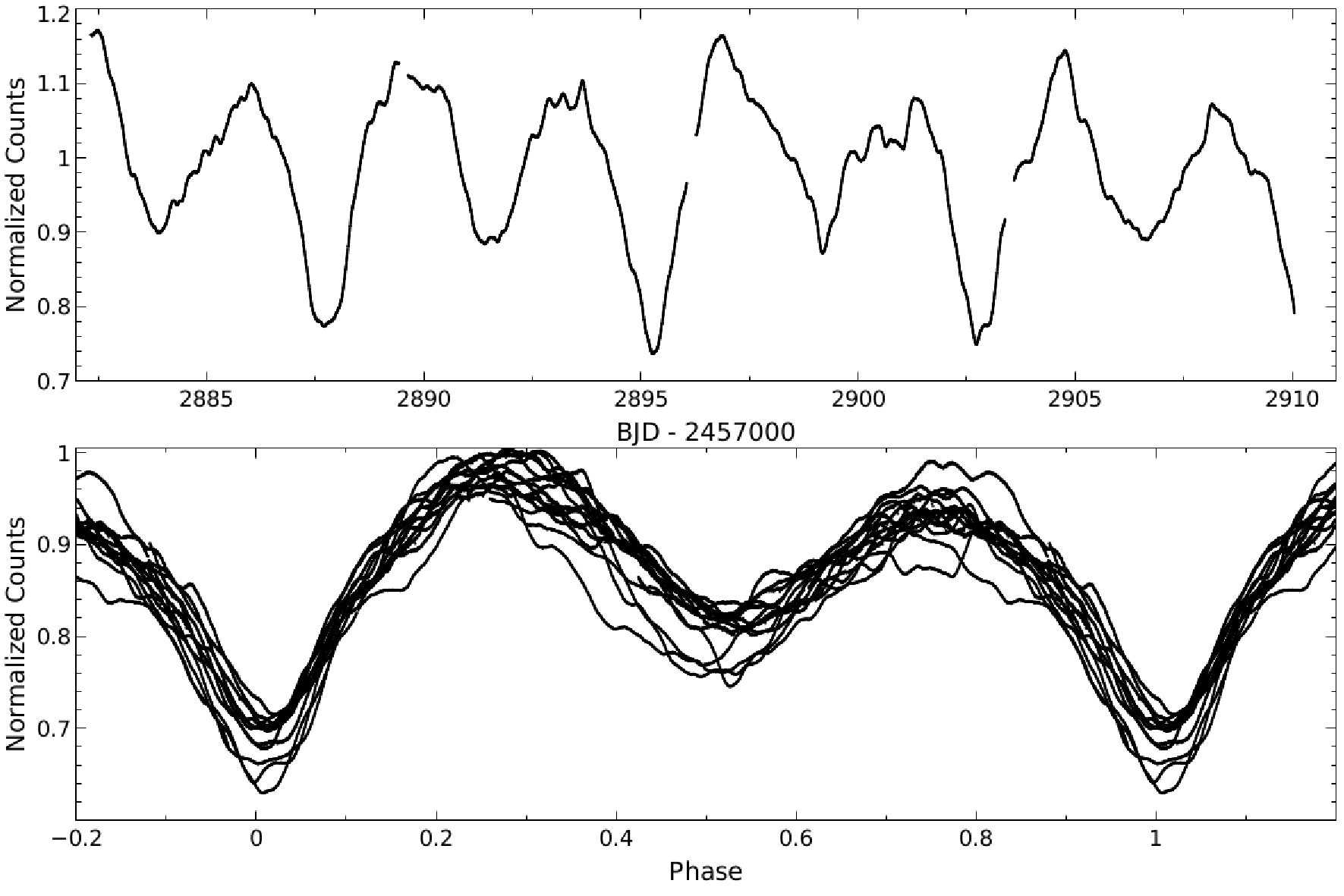}
  \caption{The top panel shows photometry of HD~5501 from one particular pointing of the {\it TESS}
    spacecraft. The photometric error bars are of similar size to the points used in the plot.  The
    horizontal axis is the barycentre-corrected Julian date (BJD) minus 2457000.  Note
    the dramatic changes from cycle to cycle in the shape and depth of the primary and secondary
    eclipses. The bottom panel shows phased HD~5501 photometry from all four {\it TESS} pointings available at
    the time of writing.  The photometry was phased using the revised ephemeris (Equation 2).}
  \label{fig:TESS1}
\end{figure}

HD~5501 has been observed by the {\it TESS} spacecraft \citep{Ricker2015} during four separate pointings. Those
observations more clearly show the salient features of the light curve of HD~5501 than the ground-based
photometry presented in previous sections.  In particular  Figure \ref{fig:TESS1} shows,
in the top panel, a single data set which, except for three short gaps, is essentially continuous over
nearly four periods.  The bottom panel shows all of the available {\it TESS} datasets phased with the
revised ephemeris (Equation 2). Each of those datasets have been normalised to unity at their highest
points.  In the upper panel of that Figure, the first eclipse is a secondary eclipse, and the second is
a primary eclipse and so on.  It can be seen that both the primary and secondary eclipses show
remarkable variability from one orbit to the next.  That variability extends to the depth, shape, width
and even (as is more clear in the lower panel) the timing of the eclipse.  Another interesting feature
of the light curve is the brightness level between eclipses.  Note that the binary {\it usually} attains
its brightest point between the primary and the secondary eclipse (phase $ \approx 0.25$), as compared
to the symmetrical interval between the secondary and the primary (phase $\approx 0.75$).  Occasionally
that is not the case, as can be seen in the upper panel: the two inter-eclipse intervals flanking
the second secondary eclipse are of nearly equal brightness.  We
will discuss the possible cause of this asymmetry in the light curve and its occasional absence in
\S \ref{sec:lcurve}. 

The dramatic variation in the light curve from cycle to cycle must imply that one or more of the components
involved in the eclipses changes size and/or shape in a complex non-periodic and possibly chaotic way.
This matter will be addressed in more detail in \S \ref{sec:lcurve} , but it seems unlikely that
the {\it stellar} components could vary in shape and size sufficiently in order to account for the light
curve irregularity.  This suggests that a non-stellar component such as an accretion disc or torus is the most likely culprit.

The highly variable nature of the HD~5501 lightcurve is reminiscent of the W Serpentis class of interacting binary stars \citep[see][]{Gies2025, Pavlovski2006}.  Stars of that class are probably in the stage of intense mass transfer and their periods are {\it increasing}, contrary to what we observe for HD~5501.  It appears that the brightest flux source in the W Serpentis binaries is the accretion torus, with the irregular lightcurve brightness changes arising from temperature and/or density inhomogeneities in the torus. As will be seen below, the brightest component in the HD~5501 system is probably the visible primary star.

The highly variable nature of the HD~5501 lightcurve raises the question of whether that variation might arise
from chaotic dynamics in the binary system.  While a detailed dynamical model of this system is beyond the
scope of this paper, we may approach this question by examining the nature of a ``reconstructed'' phase
diagram for the system.  Most readers will be familiar with the concept of phase diagrams and
  limit cycles from the study of differential equations.  A good review may be found at \citet{Limit2024}.  For instance, the phase diagram of a second-order
  differential equation is formed by plotting the dependent variable against its derivative.  If the solutions to that differential equation are stable, trajectories in the phase diagram will tend
  to approach a limit cycle with time.  Unstable solutions will diverge from a limit cycle.  Differential equations that show chaotic behavior will either show no limit cycle or one that changes with the independent variable (usually time).  Familiar examples are the
  Duffing oscillator \citep{Kanamaru2008} and the Lorenz system \citep{Lorenz1963}.

\citet{Roux83} demonstrated that it is possible to ``reconstruct'' a phase diagram for a dynamical system
from a time series.  That reconstructed phase diagram will share important properties with the actual phase
diagram of the system. In particular,
lack of a limit cycle in the phase diagram indicates the possible presence of chaos in the system.  Actual
proof of the presence of chaos necessitates calculation of at least the first Lyapunov exponent \citep[cf.][]{Miller64},
but for a binary system that requires a continuous time series over scores or even hundreds of orbital periods.  Such data sets are not generally available.

\begin{figure}
  \includegraphics[width=3.3in]{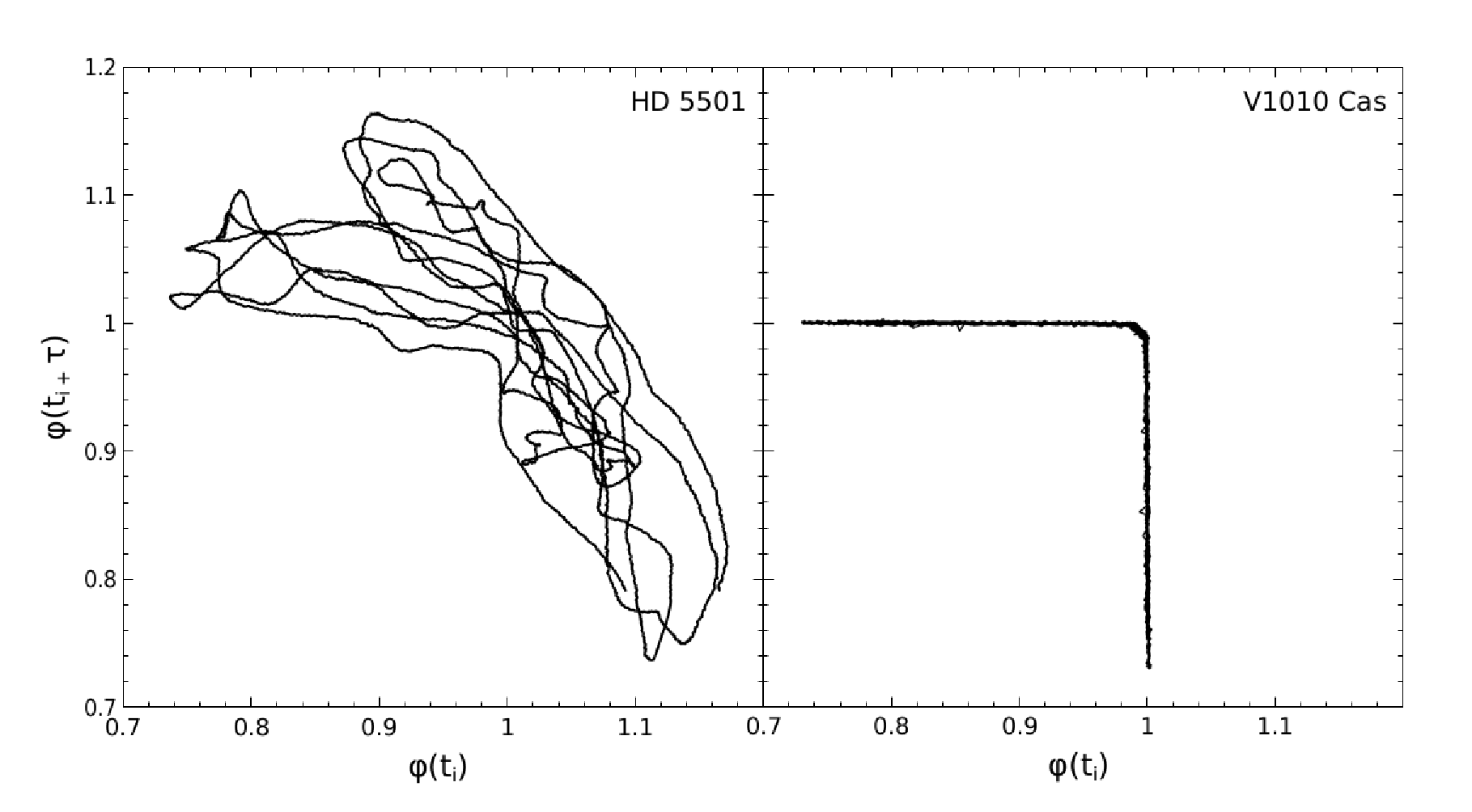}
  \caption{Reconstructed phase diagrams for HD~5501 and V1010 Cas, a ``normal'' eclipsing binary,  based on
    {\it TESS} lightcurves and the formalism
    of \citet{Roux83} (see text). Note the lack of a coherent attractor (limit cycle) in the phase diagram for HD~5501 in contrast to
  that of V1010 Cas.}
  \label{fig:chaos}
\end{figure}

To ``reconstruct'' a phase diagram for HD~5501 we need a photometric time series with low noise levels
which
is as long as possible with no gaps or very short gaps and a uniform time spacing (cadence).  The {\it TESS} data
illustrated in the top panel of Figure \ref{fig:TESS1} is adequate for this purpose.  It stretches over nearly four orbital
periods with three short gaps, and the photometry has very high precision.  Those gaps were filled
with linearly interpolated points.  In the formalism of \citet{Roux83} the time series, $\phi(t)$, is
used to ``reconstruct'' a two-dimensional phase diagram curve by forming ``tuples'' $[\phi(t_i),\phi(t_i + \tau)]$ where $\tau$ is an arbitrarily chosen time delay, usually taken as a fraction of the period of the system.  We chose $\tau = 0.75P_{\rm orbital}$, but the results are similar for any reasonable choice of $\tau$.  The result is shown
in Figure \ref{fig:chaos} along with a similarly calculated phase diagram for the ``normal'' eclipsing binary
V1010 Cas which {\it TESS} observed simultaneously with HD~5501 (2022 Nov, both in sector 58).

The difference between the phase diagrams for the two binary systems is stark.
V1010 Cas shows a well-defined limit cycle (the flipped ``$\Gamma$'') in its
phase
diagram which it repeats cycle after cycle with little or no variation.  In
contrast, the phase diagram for HD~5501 does not have a well-defined limit
cycle, but rather traces a path which is non-repeating although bounded.
While this does not prove the existence of dynamical chaos in the HD~5501
system, it does show that some physical component involved in the eclipses
varies in size or shape in an erratic and probably unpredictable way.
Interestingly, the phase diagram path for HD~5501 appears to bifurcate in the
upper left-hand
quadrant of the diagram.  This may be related to the possibility that the
secondary eclipse appears to alternate between a rounded and a more pointed
shape (see Figure \ref{fig:TESS1}).  Further investigation of this phenomenon may lead to some
insights into the dynamics of the system.

\subsection{Spectral Classification}
\label{sec:spt}

We originally identified HD~5501 as a late B-type shell star.  Unusually for
classical A-type shell stars \citep[see][]{gray09}, HD~5501 shows a variable shell
spectrum.  Classifying the spectra of HD~5501 on the MK system at different
phases can yield insights not only into the nature of the circumbinary shell,
but also the underlying star.

\citet{gray09} have described the classification of classical shell stars in
detail.  To summarise, late B and early A-type shell stars are characterised by
certain spectral features (such as the wings of the Balmer lines) that
are characteristic of dwarf, subgiant or giant stars. However, other spectral
features, in particular lines of Fe~{\sc ii} and Ti~{\sc ii} (the same lines that are
used in the luminosity classification of A-type stars) are as strong or
even stronger than those found in A-type supergiants.  Those lines  -- the
so-called ``shell lines'' -- including
Fe~{\sc ii} $\lambda 4233$, the lines of the ``Ti~{\sc ii}, Fe~{\sc ii} forest'' near
$\lambda 4500$ and, in particular, the lines of Fe~{\sc ii} multiplet 42
($\lambda\lambda$ 4924, 5018, 5169) are significantly enhanced in the spectra
of A-type shell stars.  Those lines, which arise from low-lying metastable
levels in the ion, are formed in a lower density circumstellar
shell or disc.  The goal of spectral classification is not only to detect
the presence of shell features in the spectrum, but to try to deduce the
spectral type of the underlying star.  The latter can be accomplished, at least
in the less extreme cases, by considering
spectral features that are primarily photospheric in origin, including the
wings of the higher Balmer lines (if unaffected by emission), the He~{\sc i} lines,
and high-excitation metallic lines, such as Mg~{\sc ii} $\lambda 4481$.  

\begin{figure} 
\includegraphics[width=3.2in,angle=0]{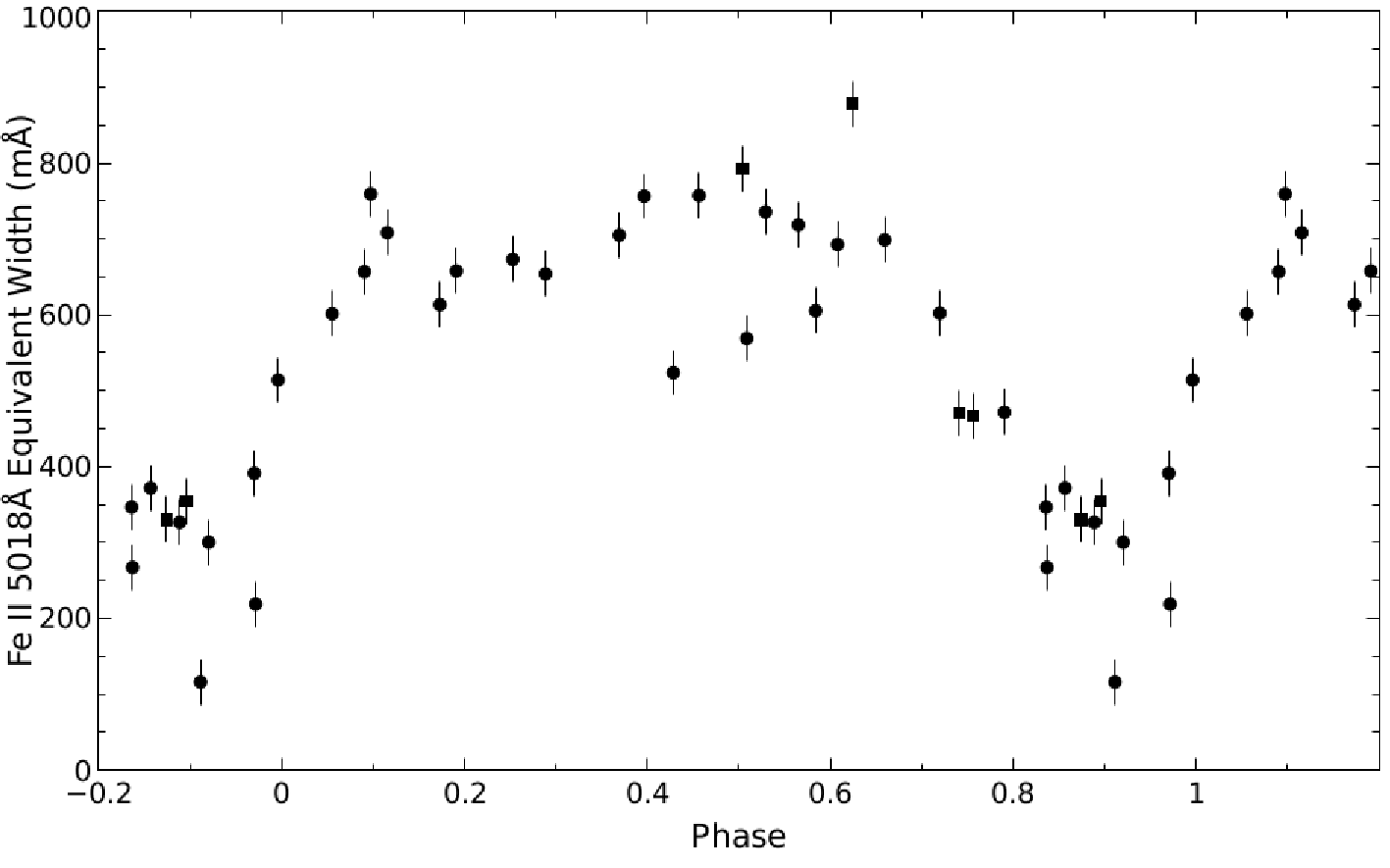} 
\caption{The equivalent width of the Fe~{\sc ii} ``shell'' line, $\lambda 5018$
  as a function of orbital phase.  Data for this figure were measured from
  VATT (squares) and SMM (circles) spectra.  The error bars were estimated
  from multiple measures.  Note that the equivalent width comes to a minimum
  just before the primary eclipse, and shows a broad maximum near the secondary
  eclipse.  The variation at a given phase is considerably larger than the
  measurement error.}
\label{fig:EQW}
\end{figure}

Table \ref{tab:mktype} records the variable spectral type of HD~5501 as a
function of phase.  The spectral types in that Table record
the average spectral type of the star at the various phases, including the
ranges observed for two spectral features in particular -- the spectral type
of the Ca~{\sc ii} K-line, and the strength of the shell features, expressed as
an ``index'' based on the full range of shell feature strengths exhibited
by HD~5501 in the spectra available for this study.  That index runs from 0 (the weakest) to
5 (the strongest), where 0 represents shell features similar in strength to those found in
normal B9 II stars (as seen in the MK standard stars).  That range corresponds
to a factor of about 2 in the equivalent width of the various shell
lines (see Figure \ref{fig:EQW}).  As an example
(see the table) at phases near the primary eclipse (the first entry in the table), the
K-line spectral type ranges between A1 and A1.5.  The strength of the shell ranges
from 0 (that is, no enhancement of the shell features at all above that seen in
standard stars) to 1 on that scale.  The table also gives more detailed comments
about other spectral features, in particular the profiles of the hydrogen lines.
The spectral types in Table \ref{tab:mktype} were determined from spectra obtained
at DSO and the Vatican Observatory, both with resolutions in the blue-violet of about
3000.
{\small
\begin{table*}
  \caption{Spectral Classification of HD~5501 as a Function of Orbital Phase}
  \label{tab:mktype}
  \resizebox{\textwidth}{!}{%
  \begin{tabular}{lll}
    \hline
    \hline
    Phase & Spectral Type & Comments (based on comparison with an MK standard at B9 II) \\
    \hline
    0.90 -- 0.10  &    B9/A0 II-: kA1-A1.5 shell: 0-1  &    Balmer cores shallow, metallic lines broad. Shell lines not enhanced at primary eclipse\\
    & & \\
    0.10 -- 0.25 &    B9/A0 II:  kA1-A1.5 shell: 2-3  &    Balmer cores slightly shallow to normal.\\
    & & \\
    0.25 -- 0.40 &   B9/A0 II: kA0.5-1.5  shell: 1-3  &   Balmer cores are generally shallow to very shallow.  Most metallic lines
    appear double near phase 0.4.\\
    & & \\
    0.40 -- 0.60 &   B9/A0 II: kA0.5-1.5 shell: 1-4   &   Balmer cores are generally shallow, but in 2005, one spectrum showed cores deeper than the standard.\\
    &                                   &   Shell lines and the Ca~{\sc ii} K-line may vary considerably in strength over time-scales $\ge P_{\rm orbit}$\\
    0.60 -- 0.80 &   B9/A0 II: kA1-A2.5 shell: 1-5  &     Balmer cores have generally shallow to normal depths, although one spectrum in 2023 showed cores \\
                 &                                 &      slightly deeper than the standard.  K-line may be the strongest at this phase, but highly variable\\
    &			           &    on time-scales $\ge P_{\rm orbit}$; same for the shell lines.\\
    0.80 -- 0.90 &   B9/A0 II-: kA1 shell: 0-2      &     The spectrum appears closest to "normal" at this phase.  In higher resolution spectra the metallic\\
                 &                                 &     lines achieve their narrowest profile.  Shell lines generally show little to no enhancement.\\
    \hline
  \end{tabular}}
\end{table*}    
}

Note that the underlying spectral type of the primary star (we cannot see any spectral features due to the secondary) is
about B9 or A0.  While the K-line type varies between A0.5 and A2.5, examination of the profile of the line in medium-resolution
spectra reveals the presence of an interstellar or circumbinary component.  The intrinsic K-line strength implies a spectral
type of about A0, but the helium lines and the hydrogen-line profiles are more consistent with B9 or B9.5 unless the primary is
a supergiant. Hence, we write the spectral type as B9/A0 to reflect that ambiguity. The Balmer line strengths and profiles are clearly affected by emission and vary in line with the highly variable emission at H~$\alpha$.  The cores are
usually quite shallow, and thus are superficially similar to those of
an A0 Ib supergiant, which explains the Simbad spectral type (A0 Ib) but the
wings of the lines imply a less luminous type.  The profiles of even the
higher Balmer lines are often asymmetrical and quite often
the photospheric wings are completely obscured.  Because the luminosity
classification near B9/A0 depends strongly on the profiles of
the Balmer lines, the luminosity type recorded in Table \ref{tab:mktype} is
highly uncertain, and that uncertainty is indicated with a colon.  While the
recorded luminosity class II: implies a fairly luminous star with
$\log g \sim 2.0$, the star may be less luminous than that, and we will present
photometric and other evidence below that the primary is probably
a ``giant'' (III) instead of a ``bright giant'' (II).

The metallic-line shell features are generally weakest near primary eclipse
(phase = 0) and strongest
during and just after the secondary eclipse (phases 0.4 -- 0.8).  This behavior
of the shell features is confirmed by measurement of the equivalent widths of
the Fe~{\sc ii} $\lambda 5018$ line (see Figure \ref{fig:EQW}), which reveals that
the minimum in the strength of the shell lines occurs just before
the primary eclipse.  Indeed, the
spectrum of HD~5501 appears most 'normal' (i.e. most similar to that of the
B9 II- MK standard) at precisely the phase where the shell features come
to a minimum.  It seems most likely that these shell lines are formed in
a circumbinary disc or ``shell'' and that we are viewing the binary through
that shell/disc.  The fact that the shell lines vary substantially in
strength over
the orbital period suggests that that disc is non-homogeneous; it may be that
at the phases when the line strengths are at a maximum (phases 0.2 -- 0.7) we
are looking along or through a relatively dense outflow from the binary.

\begin{figure} 
\includegraphics[width=3.2in]{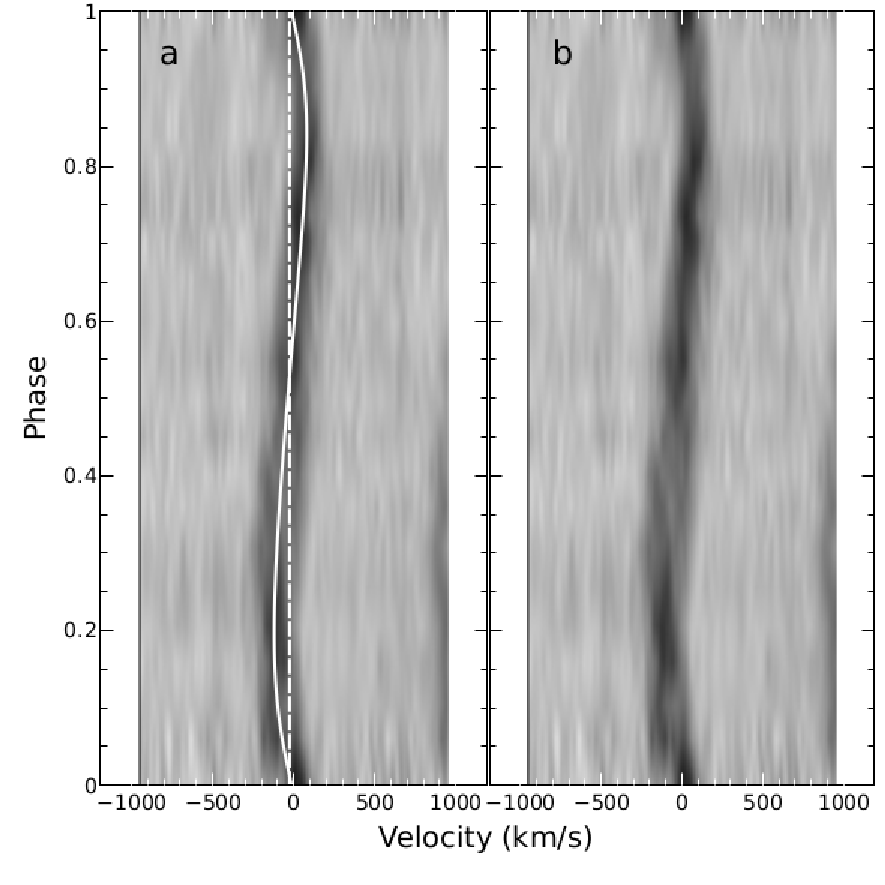} 
\caption{A grayscale rendition of the variation of the Si~{\sc ii} $\lambda$6347
  line in velocity space as a function of phase.  In panel (a) the
  line is shown with the radial velocity solution (solid white line) based on
  that line and the flanking Si~{\sc ii} $\lambda$6371 line (see \S \ref{sec:RV})
  along with the systemic velocity (white dashed line).  Panel (b) shows
  the same rendition, but without the velocity lines.  This figure illustrates
  some of the complexity of the line profile variations seen in this line and
  in others, further complicated by the orbit-to-orbit variations in the
  line profile.  This rendition is derived from spectra taken over a number of
  orbits.  Note that the profile broadens and even doubles between phases 0.3
  and 0.4.  During some orbits the doubling is very clear, in others the profile
simply broadens.  This figure is based on continuum-normalized spectra.  A continuum level of 0.83 corresponds to black.}
\label{fig:lines}
\end{figure}

The FWHM of most metallic lines in
the spectrum varies by about a factor of two during the orbital period, which
makes it impossible to determine an intrinsic $v\sin i$ for the primary (see
Figure \ref{fig:lines}), at least with the spectral material available for this study.
Indeed, near phase = 0.4 most metallic lines show a double profile, but the
implied radial velocities do not correspond to the inferred velocities for
the secondary star (see \S \ref{sec:RV}).  The FWHM is, again,
at a minimum just before primary eclipse.

Based on the spectral type, and considering its uncertainties, we suggest
$T_{\rm eff}({\rm Primary}) \approx 9750 \pm 500$K.

\subsection{Radial Velocity Solution}
\label{sec:RV}

\begin{figure}
  \includegraphics[width=3.2in]{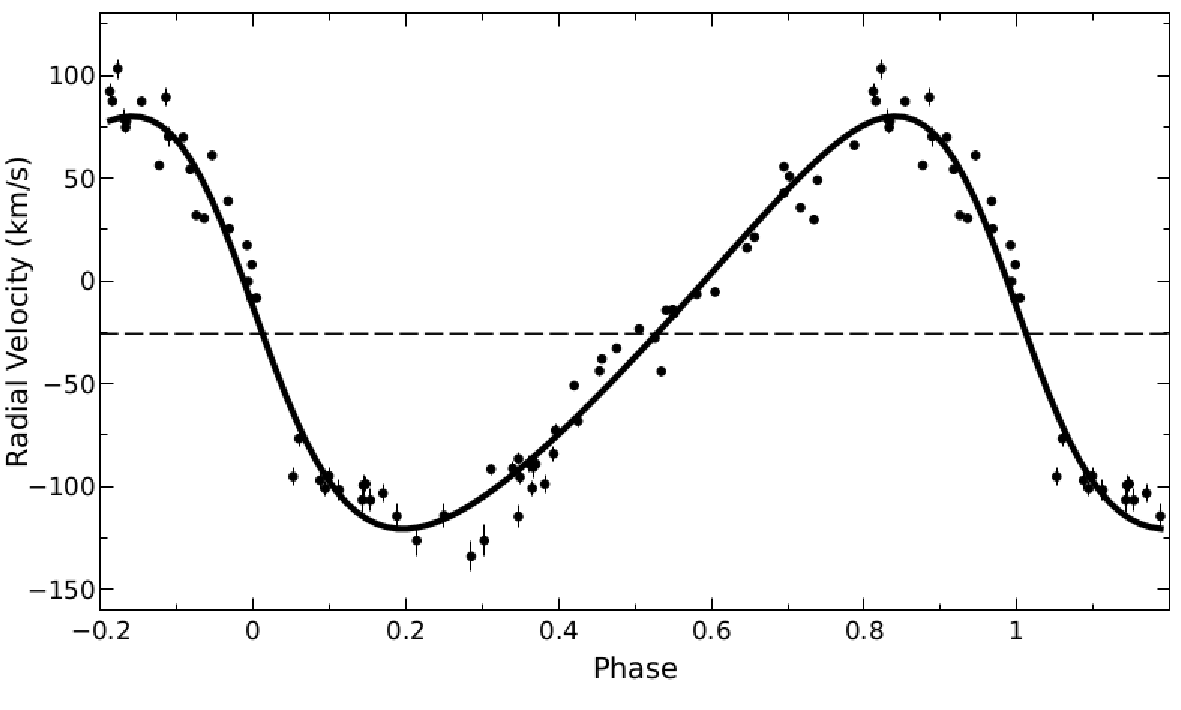}
  \caption{The radial velocity curve for HD~5501, based on observations of
    the Si~{\sc ii} $\lambda\lambda$6347 and 6371 lines (dots).  The radial velocity
    solution, discussed in the text and tabulated in Table \ref{tab:binsolve}
    is shown as a solid dark line.  The system velocity ($-25.58$ km s$^{-1}$)
    is shown as a dashed line. The horizontal axis is the light curve phase
    with the primary eclipse occurring at phase = 0.  The error bars for
    each point are based on the uncertainties of the parameters of the
    parabolic fit to the peak of the cross-correlation curve.  
    Observers who contributed data for this figure are SMM, HSO, WCO and
    NSQ.  See Table \ref{tab:spectra}.}
  \label{fig:binsolve}
\end{figure}

HD~5501 appears to be a single-lined spectroscopic binary, as careful
inspection of the medium-resolution spectra we have available for this study
(from sources SMM, HSO, and NSQ -- see Table \ref{tab:spectra}) has not yet
shown any evidence for the companion.  This may change with the acquisition
of higher signal-to-noise spectra.  Having said that, all spectral lines
deriving from metallic species (with the exception of the Na~{\sc i} D doublet) show
continuously varying profiles, including line ``doubling'' at phases between
0.3 to 0.4 (see Figure \ref{fig:lines}).  Lines of Ti~{\sc ii} and Fe~{\sc ii} show
more complex variable profiles; this is undoubtedly due to the presence of
circumbinary material.  That complex behavior makes those lines unsuitable
for the determination of the radial velocity curve.  The Ca~{\sc ii} K-line is
partly photospheric, but is blended with a constant velocity component which
is either interstellar or circumbinary.  Both lines of the
Na~{\sc i} D doublet show only a single constant velocity component which
likely is entirely interstellar and/or circumbinary.  Indeed, most parts of
the optical spectrum of HD~5501 are unsuitable for radial-velocity determination
via the cross-correlation technique because of similar contamination problems.
Fortunately, two high-excitation (8 eV) lines of Si~{\sc ii} in the red part of
the spectrum ($\lambda\lambda$ 6347, 6371), uncompromised by nearby lines
contaminated by circumbinary material or telluric absorption, appear to be
adequate for the purpose.  Because of their high excitation, those lines are
likely to be purely photospheric and without a circumbinary component.
Those lines were measured via cross-correlation \citep[using the crosscorrRV tool from the PyAstronomy package][]{pya} with
a synthetic spectrum of HD~5501~A (see \S \ref{sec:parameters}) computed
with the spectral synthesis program \textsc{spectrum} \citep{gray94}.  The
radial velocity measurements along with the radial-velocity solution
\citep[computed with the {\sc binarystarsolver} package][]{barton2020,milson20}
are shown in
Figure \ref{fig:binsolve}.  The radial velocity solution is tablulated in
Table \ref{tab:binsolve}.

{\small
\begin{table}
  \caption{Radial Velocity Solution for HD~5501}
  \label{tab:binsolve}
  \resizebox{\columnwidth}{!}{%
  \begin{tabular}{lll}
    \hline
    \hline
    Parameter & Value & Comments \\
    \hline
    $\gamma$  & $-25.58 \pm 0.12$ km s$^{-1}$ & systemic velocity\\
    $K$       & $100.34 \pm 0.17$ km s$^{-1}$ & velocity amplitude\\
    $\omega$  & $76.94 \pm 0.41$ degrees & argument of the periastron\\
    $e$       & $0.2365 \pm 0.0017$ & eccentricity \\
    $T_0$     & $60256.600 \pm 0.008$ & epoch of periastron (JD-2400000)\\
    $P_{\rm orbital}$ & 7.531151 days & orbital period (fixed)\\
    $a\sin(i)$ & $10.097 \pm 0.018$ Gm & projected semi-major axis\\
    $f(M)$ & $0.72322 \pm 0.00381$ & mass function\\
    \hline
    \end{tabular}}
\end{table}
}

According to the solution in Table \ref{tab:binsolve} the quadrature phases (i.e. the phases at which the binary axis is perpendicular to the line of sight, as measured from the primary eclipse) are at approximately 0.20 and 0.84. The periastron occurs at phase $0.977 \pm 0.018$ (again measured from primary eclipse) and so occurs shortly before primary eclipse minimum light.  These values, of course, apply only to the epoch during which the spectra used for the radial velocity solution were obtained, and we might expect them to change as the orbit evolves.  

A notable feature of the radial velocity solution is the moderately high
eccentricity ($e \sim 0.24$) of the orbit.  \citet{Mayer2006} caution that the eccentricities for early-type (massive) binaries may be spurious, and so the error bars for that parameter in Table \ref{tab:binsolve} may be unreasonably small.  Indeed, it is very difficult to explain the origin of such a high eccentricity (see discussions in sections \ref{sec:eccentricity} and \ref{sec:MESAbin}).  Having said that, the light curve is consistent with a somewhat elliptical orbit, as the centroid of the secondary eclipse is not exactly at phase 0.5 but closer to 0.55 (see \S \ref{sec:lcurve}).  As we shall see in Sections
\ref{sec:parameters} and \ref{sec:MESAbin} even a slightly eccentric orbit has
important implications for the evolution of the system.

Given a mass for the primary star, the mass function $f(M)$ (see Table \ref{tab:binsolve}) may be used to deduce the mass of the secondary, assuming a value for the inclination $i$.  A direct estimate of the mass ratio of the system may be deduced from the $v\sin i$ of the visible component and the velocity amplitude $K$ \citep[see Eq 3.9 in][]{Eggleton2006}.  Unfortunately, in the case of HD~5501, we are stymied by the line profile changes, even in the Si II doublet used for the radial velocity measurements (see Figure \ref{fig:lines}), that foil our efforts to determine the rotational velocity.  The cause of those line profile variations are not known, but since those high excitation lines should be formed primarily in the photosphere of the primary, we presume that they arise from velocity fields on the surface of the primary and/or possibly the inner part of a mass outflow (see \S \ref{sec:mhalpha}).  That question may only be resolved with higher resolution spectral data than available for this study.  

Another important caveat is the accuracy of the value of the
  argument of the periastron given by the radial velocity solution ($\omega = 77^\circ$).  Many
  interacting binaries which show distorted or variable line profiles also have $\omega$
  values within the first quadrant ($0^\circ < \omega < 90^\circ$).  This
  statistical preference is called the Barr effect \citep{barr1908}. \citet{abt2009} studied this effect and found that the majority of binary systems that exhibit the Barr effect have early B (B0 -- B3) primaries, although a small proportion of systems with late B- and A-type primaries may also show this effect.  Since the primary of the HD~5501 system is a B9/A0 star, the value of $\omega$ that we
  have derived (which lies in the first quadrant) may be suspect.

It is of interest that the Fe~{\sc ii} shell lines do show radial velocity variations
in phase with the Si~{\sc ii} lines, with only a slightly lower amplitude ($\sim
96.0 \pm 0.3$ km s$^{-1}$).  However, the systemic velocity of those lines
is significantly more negative ($-44.9 \pm 0.2$ km s$^{-1}$).  This suggests
that the Fe~{\sc ii} shell lines are formed in the inner part of the circumbinary
shell/disc which is expanding outward, perhaps as a consequence of continual or
episodic mass loss from the system. This is consistent with our observation in
\S \ref{sec:spt} that the varying equivalent width of these lines implies
a non-homogeneous circumbinary disc/shell.  On the other hand, the Ca~{\sc ii}
K-line shows two absorption components.  One component varies in radial
velocity in phase
with the Si~{\sc ii} lines, but the other component shows a constant velocity
close to the systemic velocity ($-25.6$ km s$^{-1}$).  That component
is presumably formed in the outer cooler part of the circumbinary shell/disc.
Finally, the Na D lines, which are typically very weak in the spectrum of a late
B-type star and thus must form exclusively in the shell/disc, each show a
single strong absorption component with a velocity close to the systemic
velocity. Those lines must also form in the outer cooler part of the shell/disc.

\subsection{Reddening and the Stellar Energy Distribution}
\label{sec:SED}

Since no sign of the secondary star can be seen in our spectra, we presume that
either the secondary is too faint or that it is surrounded by an optically
thick, possibly dusty, accretion disc or torus similar to that found in the
$\beta$ Lyrae system \citep[see][]{hubeny91}.  For reasonable values of the mass of the primary (a few solar masses or greater -- see \S \ref{sec:parameters}) the mass function from the radial velocity solution (see Table \ref{tab:binsolve}) implies that the secondary star has a lower mass than the primary.  If the primary fills or nearly fills its Roche lobe, this means that during the secondary eclipse, when the visible primary star eclipses the secondary component, the observed flux should come principally from the primary.  This enables us to get a handle on the reddening of the system and the luminosity of the primary star.  The large gray star symbols  in panels U, B, V and R$_C$ of Figure \ref{fig:UBVR}, which are located at the minimum brightness shown by the system during the secondary eclipse, constitute our best determination of the reddened intrinsic $U$, $B$, $V$ and $R_C$ magnitudes of the primary (see also the discussion in \S \ref{sec:lcurve}).  We obtain $U = 9.02 \pm 0.05$,
$B = 9.20 \pm 0.03$, $V = 8.77 \pm 0.03$ and $R_C = 8.46 \pm 0.03$ where
the errors are based on the intrinsic scatter in the light curve.  This gives,
for the primary, $(B-V) = 0.43 \pm 0.04$.

The reddening along the line of sight to HD~5501 may be obtained from the
3-D reddening map of \citet{green2019}.  The distance to HD~5501, $918.84^{+5.92}_{-5.89}$pc
\citep{gaia2022k} gives $E(g-r) = 0.19^{+0.03}_{-0.02}$
which corresponds to $E(B-V) = 0.19^{+0.03}_{-0.02}$.  Applying this colour
excess to the $(B-V)$ colour at the secondary minimum yields
$(B-V)_0 = 0.24 \pm 0.05$ for the primary.  That colour implies
$T_{\rm eff} \approx 7500$K
\citep{flower96} for the primary, and a spectral type near A8 \citep{gray09},
both of which are inconsistent with the B9/A0 spectral type
given in Table \ref{tab:mktype}.  This strongly suggests the presence
of dust in the circumbinary shell or disc.

An estimate for the total colour excess (external plus internal) of the system
may be obtained by transforming the most probable effective temperature of
the primary (given by spectral classification, see \S \ref{sec:spt}), 9750K,
to an intrinsic $(B-V)_0$ colour using the $T_{\rm eff} - (B-V)_0$ calibration of
\citet{flower96}.  This yields $E(B-V) = 0.44 \pm 0.05$, where the error
includes the uncertainty in the observed $(B-V)$ colour at the secondary eclipse
and the uncertainty in the effective temperature.

\begin{figure} 
\includegraphics[width=3.2in,angle=0]{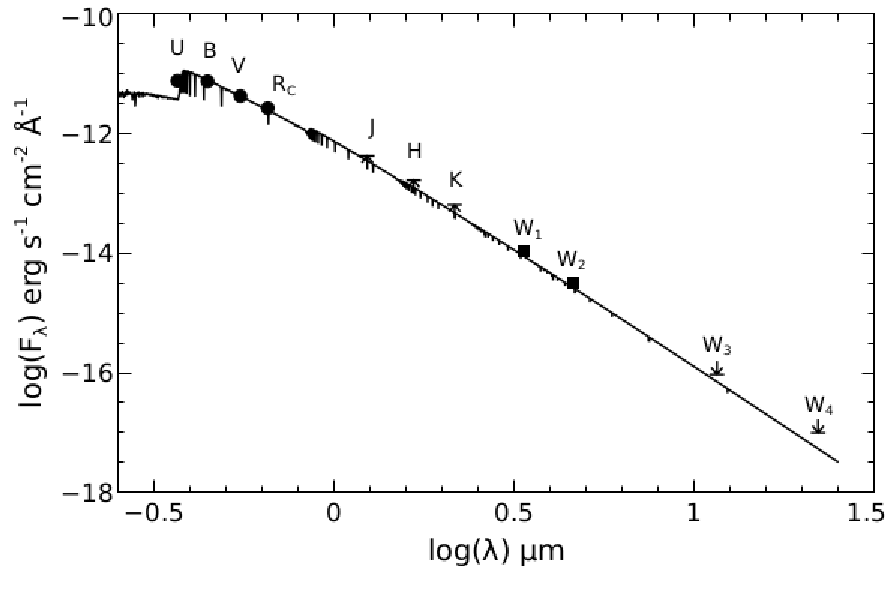} 
\caption{The stellar energy distribution for the primary star.  Fluxes
  have been dereddened assuming $T_{\rm eff} = 9750$K
  and $E(B-V) = 0.44$ for the primary (see
  text).  Points for the 2MASS $J$, $H$, and $K$ bands are upper limits,
  whereas the points for the
  {\it WISE} $W_3$ and $W_4$ fluxes are lower limits.  The model flux
  was computed with $T_{\rm eff} = 9750$K, $\log(g) = 3.0$ and [M/H] = 0
  and scaled to agree with the observed dereddened flux in the Johnson-$V$ band.
  Error bars are smaller than the symbols.  See text for more details.}
\label{fig:SED}
\end{figure}

To construct a stellar energy distribution (SED) for the primary star, we
use the typical values at the centre of the secondary eclipse of Johnson $U$,
$B$, $V$ and $R_C$ photometry derived at the beginning of this section. In
addition, HD~5501 was
observed once during the Two-micron Sky Survey \citep[2MASS][]{skrutskie06}
in the
$J$, $H$, and $K$ bands and multiple times by the {\it WISE} and {\it NEOWISE}
surveys \citep{Wright2010,Mainzer2011,Mainzer2014}
in the $W_1$, $W_2$, $W_3$, and $W_4$ infrared bands. 
There are a sufficient number of observations in the $W_1$ and $W_2$ bands to
determine typical magnitudes for those bands during the secondary eclipse.
However, the 2MASS $J$, $H$, and $K$ photometry was obtained at
${\rm phase} \sim 0.3$, which is at the brightest point in the light curve,
and so the fluxes in those bands must be treated as upper limits.  In
addition, the photometry in the {\it WISE}
$W_3$ and $W_4$ bands were only obtained between
phases 0.901 -- 0.05, that is, during the primary eclipse, and thus must
be taken as lower limits (the full light curve was observed in the $W_1$ and
$W_2$ bands because those bands could still be utilised after the end of
the cryogenic phase of the {\it WISE} spacecraft).  With those caveats, Figure \ref{fig:SED} shows our best attempt at deriving the SED for the primary star.
All flux points in the SED have been dereddened assuming $E(B-V) = 0.44$ which
corresponds to $T_{\rm eff} = 9750$K for the primary (see above).  The Figure
includes
a theoretical flux model with $T_{\rm eff} = 9750$K, $\log(g) = 3.0$,
[M/H] = 0.00 to represent the primary star.  That flux model has been scaled
to agree with the Johnson $V$-band flux.  The flux points for $J$, $H$, and
$K$ are represented as upper limits, while the flux points for $W_3$ and
$W_4$ are shown as lower limits.

Note that the $U$-band flux, which spans the Balmer discontinuity, agrees well with the model flux.  Since the size of the Balmer discontinuity in the late B and early A-type stars is exquisitely sensitive to the surface gravity, if the surface gravity of the primary star were significantly larger or smaller than the model gravity ($\log g = 3.0$) there would be a noticeable discrepancy between the observed and model $U$-band fluxes.  We shall see in \S \ref {sec:parameters} that there is a similarly good correspondence between the position of the primary in the HR diagram and a surface gravity $\log g \approx 3.0$.

HD~5501 appears to show some evidence for an infrared excess.  Since the
$J$, $H$ and $K$ flux points are upper limits, we cannot draw any conclusions
about excesses at those wavelengths.  However, the excesses at the {\it WISE}
wavelengths are small but significant.  Table \ref{tab:excess} lists the
flux excesses and their statistical significances at those bands.

\begin{table}
  \caption{{\it WISE} band infrared excesses}
  \label{tab:excess}
  \resizebox{\columnwidth}{!}{%
  \begin{tabular}{lrlllr}
    \hline
    \hline
    Band & $\lambda_{\rm eff}$ & Observed flux & Model flux & Excess flux & $\sigma$\\
         &  $\mu$m  & erg cm$^{-2}$ s$^{-1}$ \AA$^{-1}$ & erg cm$^{-2}$ s$^{-1}$ \AA$^{-1}$ & erg cm$^{-2}$ s$^{-1}$ \AA$^{-1}$ & \\
    \hline
    $W_1$ & 3.35 & $1.079(0.048) \times 10^{-14}$ & $9.040 \times 10^{-15}$ & $1.75 \times 10^{-15}$ & 3.7 \\
    $W_2$ & 4.60 & $3.240(0.060) \times 10^{-15}$ & $2.676 \times 10^{-15}$ & $5.64 \times 10^{-16}$ & 9.4 \\
    $W_3$ & 11.56 & $> 9.392(0.146) \times 10^{-17}$ & $7.102 \times 10^{-17}$ & $> 2.29 \times 10^{-17}$ & $> 15.7$ \\
    $W_4$ & 22.08 & $> 9.900(0.737) \times 10^{-18}$ & $5.374 \times 10^{-18}$ & $> 4.53 \times 10^{-18}$ & $> 6.1$ \\
    \hline
  \end{tabular}}
\end{table}

The excess fluxes recorded in Table \ref{tab:excess} depend upon an assumed effective temperature and corresponding $E(B-V)$ (see above).  Those excesses
remain robust (e.g. $\sigma > 3.0$) for a wide range of assumed effective
temperatures.  For instance, for $T_{\rm eff} = 10750$K
(1000K above the most probable temperature based on the spectral type -- see \S \ref{sec:spt}), $\sigma = 3.0$
for the  excess in the $W_1$ band.  Likewise, if we take the effective
temperature implied by the line-of-sight reddening, the significances of the
excesses are much higher than those in Table \ref{tab:excess}.  Note that the error in the adopted reddening was not used in the calculation of the excesses in Table \ref{tab:excess}.  The reason for this is that that error translates into correlated errors at each photometric band, with the result that the relative differences between the theoretical model (which is normalised to the $V$-band flux) and the photometric fluxes are hardly changed, and so the excesses in Table \ref{tab:excess} are not strongly dependent on the choice of $E(B-V)$, at least within the error bars.  However, when we calculate the luminosity of the the primary and secondary components in \S \ref{sec:parameters} the error in $E(B-V)$ must be taken into account.

  We do not have enough data to determine the origin of the infrared excesses
  noted in Table \ref{tab:excess}.  However, by analogy to the generally much
  hotter Be stars, those excesses may be produced by hydrogen free-free and
  bound-free emission from a circumbinary disc.  Alternately, the excesses
  may arise from warm dust in a circumbinary disc or shell.  A third possibility
  is the presence of a cooler third body in the system.

\subsection{The H~$\alpha$ line profile}
\label{sec:halpha}

\begin{figure}
  \includegraphics[width=3.2in]{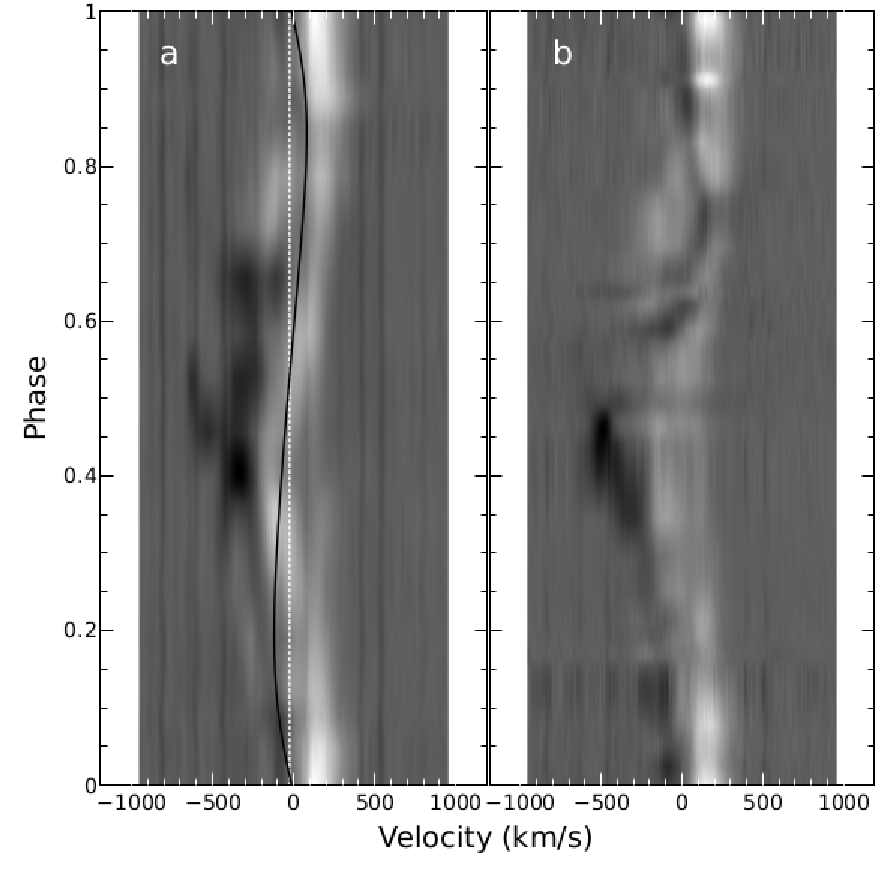}
  \caption{Grayscale renditions showing the complex variation in the
    H~$\alpha$ line of HD~5501 as a function of phase.  The horizontal
    axis is the velocity relative to the rest wavelength of H~$\alpha$ whereas
    the vertical axis is the light-curve phase.  Panel (a) is
    based on data taken during the 2023 observing season (2023 Sept -- Nov).
    The solid black curving line shows the radial velocity solution
    (Table \ref{tab:binsolve}) whereas the white dotted line indicates the
    systemic velocity. Panel (b) is based on data taken during the 2024-2025
    observing season (2024 Dec -- 2025 Feb). Note the clear differences
    between the two observing seasons, described in more detail in the text.
    These figures are based on continuum-normalized spectra.  The grayscale ranges from a continuum value of 0.50 (black) to 1.85 (white) for both panels.  The faint vertical lines in both panels are telluric lines.}
  \label{fig:halpha}
\end{figure}

HD~5501 shows strong and highly variable emission and absorption at
H~$\alpha$.  Figure \ref{fig:halpha} shows grayscale renditions of the
H~$\alpha$ line during two observing seasons, 2023 Sept -- Nov
(panel a; observer SMM) and 2024 Dec  -- 2025 Feb (panel b; observer JRF).
The horizontal axis is the velocity relative to the rest wavelength of
H~$\alpha$.  The vertical axis is the photometric phase
(primary eclipse at phase 0, secondary eclipse near phase 0.5).  Panel (a)
also shows the radial velocity solution derived in
\S \ref{sec:RV} (solid black line) as well as the systemic
velocity (dashed white line). 

\begin{figure}
  \includegraphics[width=3.2in]{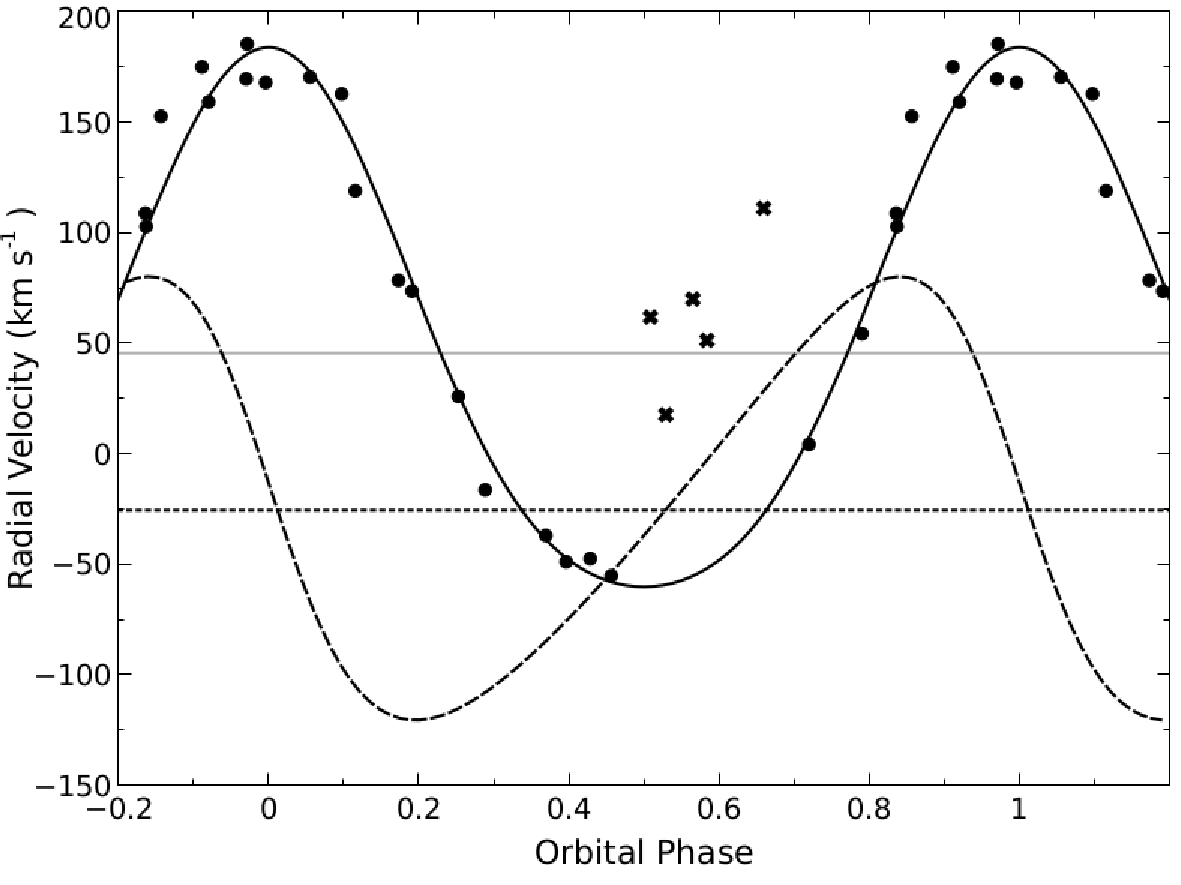}
  \caption{The points (both filled circles and crosses) represent the velocity
    of the centroid of the emission component(s) of the H~$\alpha$ profile
    during the 2023 observing season.  Note that the
    filled circles follow a sinusoidal curve with an amplitude of
    122 km s$^{-1}$, but the crosses, which represent points during the
    time that the emission profile is double peaked deviate from that curve.
    The gray horizontal line
    represents the H~$\alpha$ mean velocity.  The dashed curve is the radial
    velocity solution for the primary star derived in \S \ref{sec:RV} based
    on Si~{\sc ii} lines.  The dotted horizontal line is the systemic velocity from
    that solution.}
  \label{fig:Havel}
\end{figure}

The variation in the H~$\alpha$ profile shows considerable differences between
the two observing seasons.  During the 2023 observing system the emission
at H~$\alpha$ varied between a single-peaked and a double-peaked
profile.  Doubling took place at phases between the secondary and primary
eclipse ($\sim 0.6$ to  $\sim 0.8$) with the red peak consistently higher than
the blue peak.  During single-peaked phases the profile
often showed one or more blue-shifted absorption components, although
this varied from
orbit to orbit.  The strongest emission peak of the H~$\alpha$ profile during
the 2023 season followed a sinusoidal variation in velocity with a higher
amplitude ($\sim 122$ km s$^{-1}$) than the radial velocity curve based on
the Si~{\sc ii} lines (100 km s$^{-1}$: see Figure \ref{fig:Havel}).  During that
season the mean velocity of the H~$\alpha$ emission (45.44 km s$^{-1}$)
deviated from that of the systemic velocity based on the Si~{\sc ii} radial
velocity curve (-25.58 km s$^{-1}$) by 71 km s$^{-1}$.  It was also out of
phase with that curve.  The difference in the mean velocities implies that
most, if not all, of the H~$\alpha$ emission is not associated with either of
the stellar components of the system, but likely represents an outflow from
the system.  This will be discussed in more detail in \S \ref{sec:mhalpha}.

During the 2024-2025 observing season the predominant H~$\alpha$ emission
peak became
generally lower, broader, sometimes with structure, and on occasion showed
a flat-topped profile. During the
double-peaked phases, in contrast to the 2023 season, the blue peak was often
higher than the red peak.  However, the sinusoidal variation in the main
H~$\alpha$ emission feature is similar to the 2023 observing season, but
because the emission peak profile is somewhat broader and lower, that 
variation is not as obvious.  

\begin{figure}
  \includegraphics[width=3.2in]{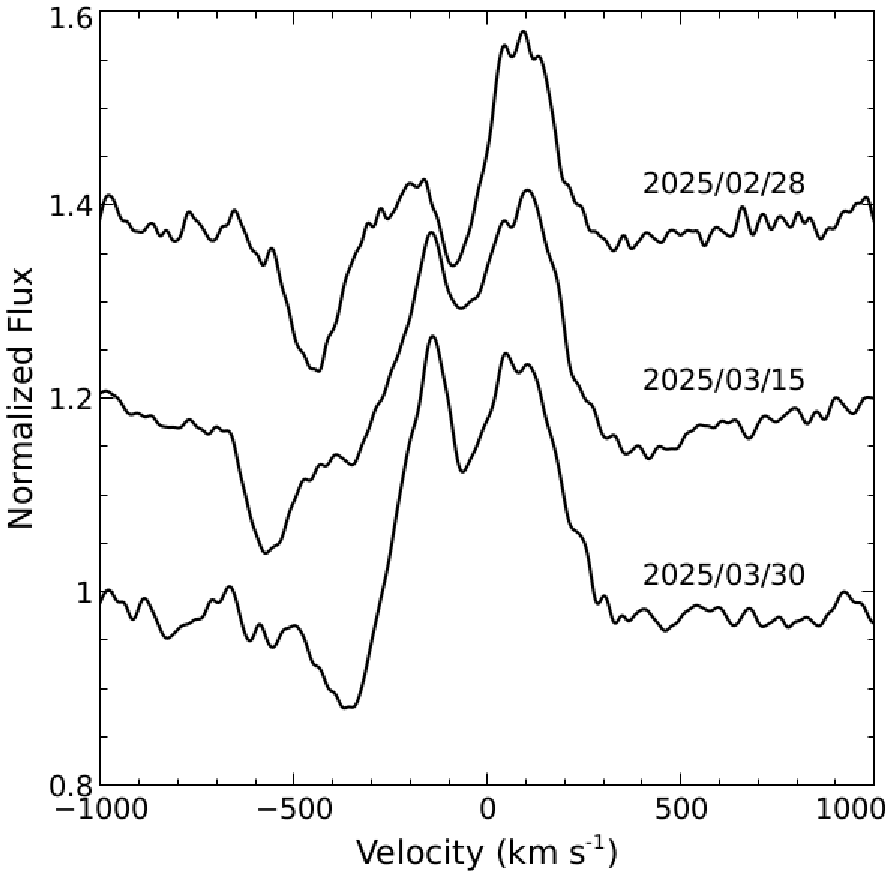}
  \caption{The H~$\alpha$ profile at nearly the same phase (0.550, 0.548 and
    0.546; top, middle and bottom respectively) observed over an interval of
    four orbits. This phase corresponds closely to centre of the secondary eclipse.
    Note the variation in the shape of the predominant emission
    component, but especially note the changes in the strengths and velocities
    of the absorption component(s). The top spectrum was observed by WCO,
    the bottom two by THO -- see Table \ref{tab:spectra}. Telluric lines have been removed from these spectra.  These spectra are continuum normalized and displaced by 0.2 continuum units for clarity.}
  \label{fig:abs}
\end{figure}

During both observing seasons the H~$\alpha$ emission and absorption components exhibited
significant variation from one orbit to the next, and the absorption components showed blue-shifted
velocities up to 500~km s$^{-1}$ with respect to the systemic velocity (see Figure
\ref{fig:abs}).  The highest
velocity absorption components are observed near to the secondary eclipse
(phases $\sim 0.4 - 0.7$).  These absorption components will be discussed
in more detail in \S \ref{sec:mhalpha}.

\subsection{Physical Parameters of the Components}
\label{sec:parameters}

To gain an understanding of the physical nature of this system, it is
necessary to estimate the masses of the components and to determine their
evolutionary states.  From the evidence considered so far, it is clear that
this is an interacting binary system, implying that the primary fills or
overfills its Roche lobe during at least some portion of its elliptical orbit.  This fact
will be useful in constraining the characteristics of the primary.  To begin
our analysis, we will assume that we are viewing the binary in the orbital
plane ($i = 90^\circ$) and then we will discuss how our analysis changes
if $i < 90^\circ$.

\begin{figure} 
\includegraphics[width=3.2in,angle=0]{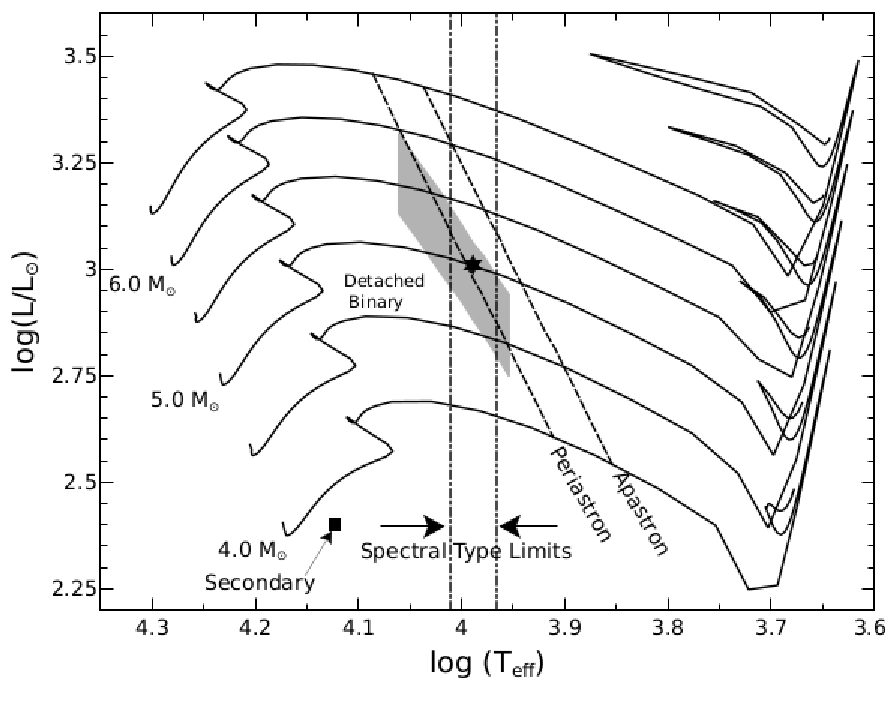} 
\caption{A theoretical Hertzsprung-Russell diagram illustrating the various
  constraints to the solution for the parameters of the primary star.  The
  evolutionary tracks (solid black lines) are based on {\sc mesa} models.  The
  zero-age main sequence (ZAMS) is represented by the beginning of the tracks
  to the left.  The
  shaded area shows the $1 \sigma$ boundaries for the photometric and
  astrometric solution discussed in the text.  The dashed lines labelled
  ``Periastron'' and ``Apastron'' represent, respectively, the loci of points
  where the primary fills its Roche lobe only at periastron and where the
  primary just fills its Roche lobe at apastron, implying that it overfills its
  Roche lobe during the remainder of the orbit.  Those lines are drawn assuming $e = 0.24$.  The vertical dot-dashed and
  dot-dot-dashed lines
  represent limits to the effective temperature of the primary imposed by
  the spectral type.  The six-pointed star indicates a ``representative''
  solution for the primary star with $M = 5M_\odot$ (see text).  The filled
  square represents the corresponding solution for the secondary star.}
\label{fig:HRD}
\end{figure}

We may place the primary component on the HR diagram by using the $V$ and
$B$ magnitudes observed during the secondary eclipse and the Gaia distance
(see discussion in
\S\ref{sec:SED}) to deduce the absolute visual magnitude $M_V$ of the primary
and thus the luminosity.  As discussed in \S\ref{sec:SED}, a significant unknown
is the total reddening, $E(B-V)$ of the system.  The line-of-sight reddening
to the star given by \citet{green2019} yields a $(B-V)_0$ colour that is too
red for the observed spectral type.  This suggests the presence of a
significant amount of dust in the system, probably in a circumbinary shell
or disc.  To proceed, it is necessary to consider models with a range of
effective temperatures and then to further constrain those models to those
consistent with the spectral type {\it and} the fact that the primary must
fill its Roche lobe during at least a portion of its orbit.  Beginning with a
specific effective temperature enables us to deduce an intrinsic $(B-V)_0$
colour from the $T_{\rm eff} - (B-V)_0$ calibration of \citet{flower96}.
That calibration also yields the bolometric correction (B.C.) of the star
from which $M_V$ and $\log(L/L_{\odot})$ may be deduced.  Errors associated
with the observed apparent magnitude during the secondary eclipse ($V = 8.77
\pm 0.03$), the colour ($(B-V) = 0.43 \pm 0.04$), the resulting colour
excess $E(B-V)$, and the Gaia distance ($918.84^{+5.92}_{-5.89}$pc) must all be
taken into account to determine the uncertainty in the resulting
$\log(L/L_{\odot})$.  For this analysis we consider models with effective
temperatures ranging from 9000K to 11500K which contains the range of
effective temperatures consistent with the spectral type as a subset
(see \S \ref{sec:spt}).  That analysis
yields the shaded polygon in Figure \ref{fig:HRD}, the vertical height of
which represents the $1 \sigma$ limits on the calculated errors for
$\log(L/L_{\odot})$ for the chosen range of effective temperatures.  Also
shown in that diagram are {\sc mesa} evolutionary models \citep[{\sc mesa} v24.08.1:][]{Paxton2011, Paxton2013, Paxton2015, Paxton2018, Paxton2019} for masses
between 4.0 and 6.5$M_\odot$.  Those models were computed assuming solar
abundances and no mass loss and were calculated up to the point of core
helium exhaustion.

What is immediately evident from this analysis is that all reasonable
solutions for the position of the primary in the HR diagram indicate that
it has undergone core hydrogen exhaustion and is currently in a phase
of rapid evolution across the Hertzsprung Gap during which the inert helium
core is contracting and the envelope expanding.  It is likely that the
primary has only recently begun to fill its Roche lobe, transforming
what was previously a detached binary into an interacting binary with mass
transfer from the primary on to the secondary.

An added complication is that the orbit of the binary is elliptical, meaning
that the primary first began filling its Roche lobe at periastron.  The
radius data from the {\sc mesa} models may be compared to the size of the Roche
Lobe to compute a locus of points in the HR diagram corresponding to the
condition that the primary fills its Roche lobe only at periastron.  That
locus of points in indicated in Figure \ref{fig:HRD} as a dashed line.
A similar locus of points may be calculated for the situation in which the
primary just fills its Roche lobe at apastron.  That line is also indicated
in Figure \ref{fig:HRD}.  In between those two lines the primary fills or
overfills its Roche lobe during the part of its orbit near periastron and
underfills during the remainder of its orbit.  This is referred to as
``phase-dependent Roche-lobe overflow''. To the right of the apastron locus
the character of the binary is completely changed due to the rapid mass
transfer. See \S \ref{sec:MESAbin} for details. Note that
  these lines were computed assuming $e = 0.24$.  If the eccentricity is more
  moderate than that, those two lines will be closer together.

The calculations for those two boundaries in the HR diagram were carried out
with the following steps.  1) Beginning with a mass $M_1$ for the primary,
the equation for the mass function may be solved iteratively for the mass
of the secondary, $M_2$.  This gives the mass ratio $q = M_2/M_1$.
2) This allows the computation of the total semi-major axis of the binary:
$a = a_1 + a_2$ where $a_1$ (assuming $i = 90^\circ$) is given by the radial
velocity solution, and then $a_2 = a_1/q$.  3) For a given mass ratio, the
parameter $R_{\rm vol}/a$ for the Roche lobe may be computed using the
Fortran code {\sc rochelobe.f90}, version 1.0 written by D. Leahy
\citep{leahy2015}.  $R_{\rm vol}$ is the radius of a sphere that has
the same volume as the Roche lobe.  The parameter $R_{\rm vol}/a$ may be
compared with the radius data in the {\sc mesa} model to determine the point in
the evolutionary track where $R_{\rm model}/a = R_{\rm vol}/a$ and thus where
the star begins to fill its Roche lobe (the periastron boundary) or where the
star just fills its Roche lobe at apastron (the apastron boundary).  While
the calculations in the
code {\sc rochelobe.f90} assume a circular orbit, the authors point out that
the equation for the Roche potential may be generalised to the case of
elliptical orbits if $p^2$, where $p = \Omega_{\rm star}/\Omega_{\rm binary}$
(the ratio of the rotational angular velocity of the primary star to the
orbital angular velocity of the binary),
is replaced in that equation with the function
\begin{equation}
  A(p,e,\nu) = \frac{p^2(1+e)^4}{(1+e\cos(\nu))^3}
\end{equation}
where $\nu$ is the true anomaly in the primary orbit.  For simplicity we
assume $p^2 = 1$ (synchronous rotation) in that equation.

The analysis above assumes that we are seeing the eclipsing binary along
the orbital plane ($i = 90^\circ$).  If $i < 90^\circ$, this leads to a larger
semi-major axis $a$ for the system, but the mass function calculation yields
a larger mass
ratio as well, with the net result that the size of the Roche lobe for the
primary hardly changes. This means that the positions of the Periastron
and Apastron loci in Figure \ref{fig:HRD} are insensitive to inclination,
even for inclinations as low as $i = 70^\circ$.  Thus the position of the
primary in this diagram cannot be used to constrain the inclination.
Analysis of the light curve will yield a stronger constraint to the
inclination, but the highly variable nature of the light curve complicates
that analysis as well.  We will assume in the discussion that follows
that $i = 90^\circ$.

The requirements that the primary must fit the Roche-lobe constraints,
be consistent with the spectral type, and fall within the shaded box (determined
by photometry and astrometric data) in Figure \ref{fig:HRD} imply that the
mass of the primary falls within the $1 \sigma$ limits
$4.6M_\odot < M_1 < 5.4M_\odot$.
With the current data, we can not further restrict that mass range.  To carry
the analysis somewhat further, however, we select a ``representative''
solution for the
primary, represented by the ``star'' in Figure \ref{fig:HRD} where
$M_1 = 5.0M_\odot$ and $T_{\rm eff} = 9750$K.  The {\sc mesa} model for that mass gives
further that $\log(L/L_\odot) = 3.01$, $\log(R/R_\odot) = 1.048$,
$\log(g) = 3.04$, with the age = $8.49 \times 10^7$yrs.  The value for
$\log(g)$ suggests that the luminosity type for the primary is closer to
III instead of II.  The mass function
(see Table \ref{tab:binsolve}) may be iterated to yield $M_2 = 3.84M_\odot$,
giving $q = 0.768$.

At that age, the secondary star has, according to a {\sc mesa} model computed for
that mass, an effective temperature of about
13265K, $\log(R/R_\odot) = 0.48$, $\log(L/L_\odot) = 2.40$ and $\log(g) = 4.07$.
  Its position on the HR diagram indicates that it is still burning hydrogen on
  the main sequence, and would be well within its Roche lobe, even at
  periastron.  The effective temperature implies a spectral type near B7 V.
  This means that in the blue-violet, in spectra of better quality and higher
  resolution than those available for this study, the secondary might best
  be detected as blends in the He~{\sc i} line profiles, or possibly by the
  presence of the C~{\sc ii} $\lambda 4267$ line.

\subsection{A Preliminary Light Curve Model}
\label{sec:lcurve}
A complete understanding of the complexity of the light curve shown by HD~5501
will need to await a full dynamical model of the system which is beyond the
scope of this paper.  However, it is
possible to take a first step toward that model beginning with the
``representative
solution'' proposed at the end of the previous section based on {\sc mesa}
evolutionary models and consistent with the radial velocity solution and the
spectral type limits.  To recall, $M_1 = 5.0M_\odot$, $T_{\rm eff} = 9750$K, and
$\log(R/R_\odot) = 1.048$ with an age of $8.49 \times 10^7$yrs.  The secondary
star has $M_2 = 3.84M_\odot$, $T_{\rm eff} = 13265$K, and $\log(R/R_\odot) = 0.48$.
We assume $i = 90^\circ$. With that assumption, the radial velocity solution
gives us $a_1 = 1.0097 \times 10^{10}{\rm m} = 14.51R_\odot$, $a_2 = a_1/q =
18.90R_\odot$, yielding $a$, the orbital semimajor axis $= 33.41R_\odot$.  The
other salient facts about the system are (i) that no evidence for the secondary
can be seen in our medium-resolution spectra, even though synthetic spectra
indicate that a bare secondary should contribute visibly to the observed
spectrum, (ii) that the orbit is elliptical with an eccentricity possibly as high as $e = 0.24$ (see discussion in \S \ref{sec:RV}), (iii)
that mass transfer is evidently occuring near periastron passage, as discussed
in \S \ref{sec:parameters} and (iv) that the light curve is slightly
asymmetrical with the part of the light curve between the
primary and secondary eclipses
{\it usually} slightly brighter than the corresponding part of the light curve
between the secondary and primary eclipses (see \S \ref{sec:chaos} and
Figure \ref{fig:TESS1}). Considerations
(i) and (iii) suggest that the secondary is hidden behind an opaque torus,
not unlike that hypothesized for the $\beta$ Lyrae system \citep{hubeny91}.
Indeed,
model light curves with a partially transparent torus or a disc with thickness
less than the diameter of the secondary completely fail to reproduce the light
curve.  Consideration (iv) suggests but does not prove that the torus has a
hotspot
at the location where a stream from $L_1$ would impact.  Another possible
cause of this asymmetry could be brightening due to phase dependent Roche-lobe
overflow at that phase.

It is important to reflect on what we are trying to fit with our model.  Examination of both the ground-based photometry (Figure \ref{fig:UBVR}) and TESS photometry (Figure \ref{fig:TESS1}) indicate that there is considerable variation in the flux at minimum light during the secondary eclipse.  If the torus is completely contained within the Roche lobe of the secondary star, the secondary eclipse should be total (assuming an inclination near $i = 90^\circ$).  We presume that this variation in the flux arises from changes in the shape, size and possibly the axial symmetry of the torus around the secondary star with the torus or parts of the torus occasionally extending beyond the Roche lobe of the secondary.  During those times the secondary eclipse would not be total.  We assume that when the flux is at the minimum observed during the secondary eclipse (corresponding to the gray ``stars'' in Figure \ref{fig:UBVR}) this  corresponds to the situation during which the torus is confined to the Roche lobe of the secondary, and thus the eclipse is total.  Our model, detailed below, assumes a total secondary eclipse.  This picture may be simplistic, as there are probably outflows from the system (see \S \ref{sec:mhalpha}) but it represents a reasonable first-order approximation to begin with.  We will evaluate in \S \ref{sec:mhalpha} the possible effect of such outflows on the light curve.

We employ the software package {\sc shellspec~v49} \citep{Budaj2004} to
derive a preliminary model for this system. While the current version of
{\sc shellspec} is
unable to handle elliptical orbits natively, the eccentricity of the orbit is not so
large that the assumption of a circular orbit is invalidating, at least to
the first order.  In this simplified circular orbit model we assume that the
primary fills its Roche lobe at all phases and that the secondary, which is
well within its
Roche lobe is spherical, and, as stated above, is surrounded by an opaque
torus which is confined within the Roche lobe of the secondary (see discussion above).  This leaves the characteristics
of that torus (outer radius, height, and surface temperature) as free
parameters.  To simplify things, we assume that the torus is a circular
slab.  Because it is opaque, the internal structure (such as the density and
temperature structure) is irrelevant.  The final model that gives a reasonable
first-order representation of the light curve 
has $r_{\rm torus} = 8.0R_\odot$, $T_{\rm torus} = 7500$K (outer surface),
and half height $h_{\rm torus} = 3.02R_\odot$, which corresponds to the radius
of the secondary star.

\begin{figure} 
\includegraphics[width=3.2in,angle=0]{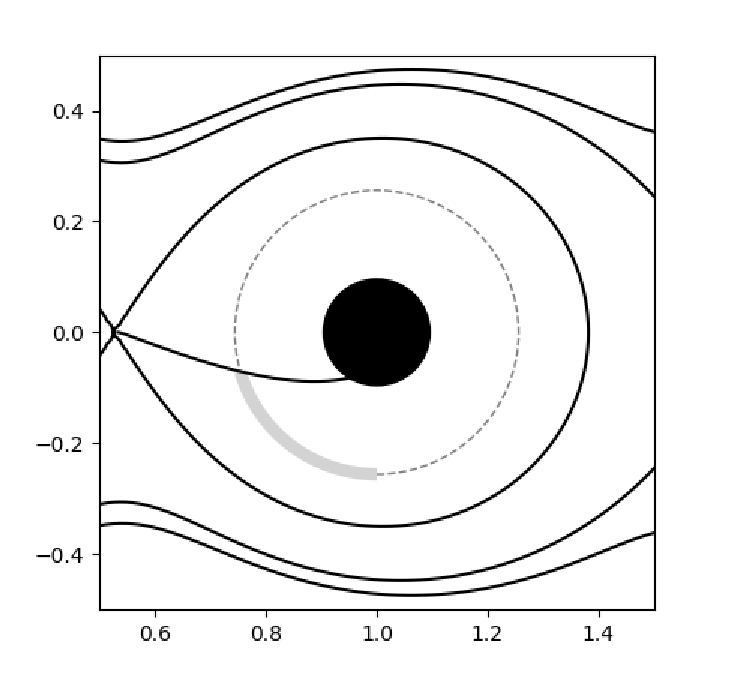} 
\caption{This figure shows the Roche potential surrounding the secondary
  star (assuming $M = 5M_\odot$ for the primary star and $q = 0.768$) at the
  beginning of Roche-lobe overflow.  The mass stream begins at rest (in the
  rotating coordinate system) at the $L_1$ Lagrange point (far left of the
  diagram) and, at the beginning of mass transfer before the formation of an
  accretion disc/torus, directly impacts the secondary star (black circle).  The
  stream line was calculated using the equations of motion for the
  restricted three-body problem (assuming initial circular orbits).
  Dimensions are in units of the semi-major axis.  The dashed circle
  surrounding the secondary star shows the outer boundary of the accretion
  disc/torus discussed in Sections \ref{sec:lcurve} and \ref{sec:mhalpha}.  The
  gray arc shows the ``hot spot'' used to model the light curve in \S \ref{sec:lcurve}.}
\label{fig:di}
\end{figure}

\begin{figure} 
\includegraphics[width=3.2in,angle=0]{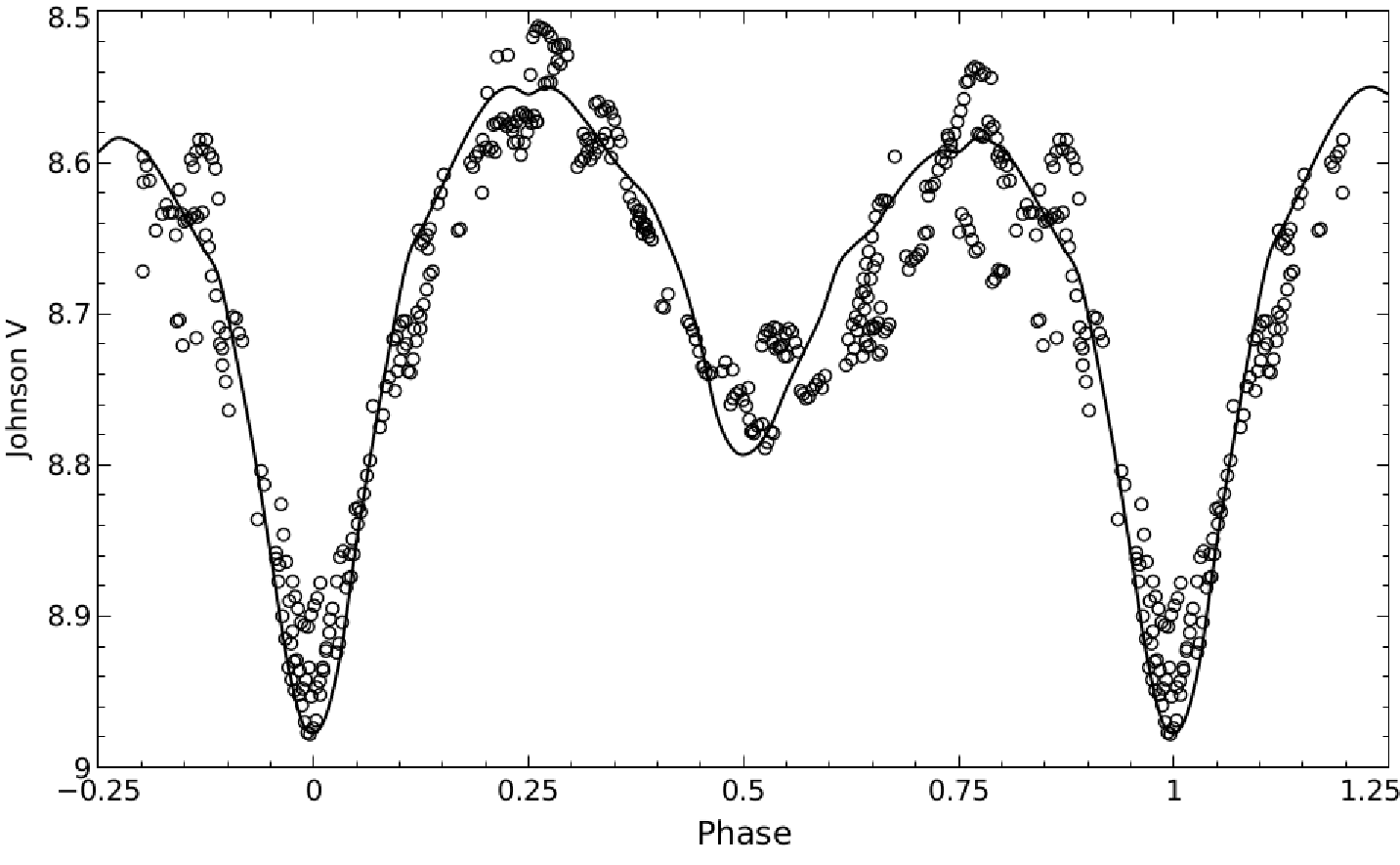} 
\caption{The open points show the Johnson $V$-band light curve for HD~5501 observed at the Dark Sky Observatory during the 2023/24 observing season.  The theoretical light curve (solid line) was computed by convolving the
    fluxes calculated by {\sc shellspec} as described in \S \ref{sec:lcurve}
    with the Johnson-$V$ passband \citep{Bessell1990}.}
\label{fig:lcmodel}
\end{figure}

The logical position to add a hotspot
  on the torus to explain the asymmetry of the light curve is at the intersection of the stream originating at $L_1$ and the boundary of the torus
  (see Fig \ref{fig:di}).  However, because of the size of that torus, a
  circular hotspot at that position does not yield the required asymmetry in the fluxes without assuming an unreasonably large spot and/or high temperature.  The asymmetry is best modeled by a hotspot elongated in the direction of orbital motion of the torus (see Fig \ref{fig:di}).  To model this, we use the ``ring'' object in {\sc shellspec} to model an arc-shaped hotspot by specifying the beginning of the arc at $196^\circ$ (measured counterclockwise from the line between the centre of the secondary and the $L_2$ point) and $270^\circ$.  The first angle corresponds to the intersection between the stream from $L_1$ and the outer boundary of the accretion torus.  The choice for the second angle is somewhat arbitrary, but helps to reproduce the light curve asymmetry quite well (see Fig \ref{fig:lcmodel}).  To achieve that agreement we set $T_{\rm hotspot} = 15000$K and take the radius of the cross-section of the hotspot arc as $0.5R_\odot$.

This first-order model fits the light curve reasonably well, considering
the approximations that were made.  The width and depth of the primary eclipse are nearly correct, but the shape of the secondary eclipse is somewhat too pointed.  Having said that, the observed shape of the secondary
eclipse is highly variable (see \S \ref{sec:chaos} and Figure \ref{fig:TESS1}).  The observed centre of the secondary eclipse occurs near to phase = 0.55, presumably a consequence of the eccentricity of the orbit, whereas the model secondary eclipse occurs at phase = 0.5 because the {\sc shellspec} model assumes a circular orbit.

A further step towards modeling the light curve might be taken with
{\sc shellspec} as elliptical orbits can be represented as a series of segments
with the instantaneous positions and velocities of the primary and secondary
input into the program.  However, an added complication arises because at
periastron passage the primary almost certainly overfills its
Roche lobe leading to enhanced mass transfer to the secondary.  When the
Roche-lobe fill factor exceeds 1.0, {\sc shellspec} models the system as a
contact
binary which results in a large distortion of the light curve.  Instead, what
likely happens at periastron passage is that the mass transfer to the
secondary is greatly enhanced, with material flowing into the Roche lobe of
the secondary, but because of the short time-scales involved ($\sim 1$ day),
not filling it.  However, during periastron the primary star overfills
its Roche lobe to the extent that mass is almost certainly lost from the
system near the $L_3$ point (see \S \ref{sec:mhalpha}).  After periastron
passage, particle stream lines indicate that the rest of the material
falls back on to either the primary or the secondary star/torus.
We expect that this rapid mass transfer during periastron passage affects
the shape, size and temperature of the torus and hypothesized hotspot,
helping to explain the unusual variability of the light curve.  The expelled
material, as will be seen in \S \ref{sec:mhalpha}, may be partially
responsible for the variable emission and absorption
features seen at H~$\alpha$.

\subsection{Time-scale for the orbital evolution}
\label{sec:orbit_evolution}

The analysis in \S \ref{sec:parameters}
indicates that the primary star in the system is undergoing rapid evolution
across the Hertzsprung gap. That rapid evolution involves a significant
increase in the radius of the star over a short period of time.  That means
that the size of the star relative to its Roche lobe is undergoing a rapid
change, which will affect the orbit of the binary via mass transfer and
possibly mass loss (see \S \ref{sec:parameters}).  Therefore, a reasonable
estimate for the
time-scale for the evolution of the orbit of the binary is the length of time
that the star takes to move between the periastron and apastron boundaries
on its evolutionary track described
in the previous section and illustrated in Figure \ref{fig:HRD}.  For the
selected $5M_\odot$ model, that works out to $\sim 90,000$ yrs.  This may be
compared to the time-scale $P/\dot{P} \sim 170,000$ yrs derived from the O-C
diagram in \S \ref{sec:orbit}.  The similarity of those two figures suggests
that this explanation for the rapidity of the orbital evolution is reasonable.

\subsection{Origin of the Eccentricity of the Orbit}
\label{sec:eccentricity}

One of the most curious features of the HD~5501 system is the computed eccentricity of the orbit, $e = 0.24$. While we have discussed the possible spuriousness of such a large eccentricity (see \S \ref{sec:RV}) it is important to examine which mechanisms might be responsible for creating a non-circular orbit.  Tidal forces associated with an eccentric orbit
should rapidly circularize the orbit upon the onset of Roche-lobe overflow.  There are, however, two
mechanisms that can act on the eccentricity of the orbit which are likely
relevant to this case.  The first concerns whether or not at the beginning of Roche lobe overflow the mass transfer
stream through the $L_1$ point will undergo a direct impact on the secondary
star as opposed to going into orbit around the secondary, possibly forming
a disc. \citet{sepinsky10} have demonstrated that direct impact accretion
can increase {\it or} decrease the orbital eccentricity depending on the
details of the accretion.  Direct impact accretion can also increase or
decrease the semi-major axis which, of course, causes an increase or
decrease of the orbital period.  Figure \ref{fig:di}, based on a direct
numerical calculation of a particle released at $L_1$ at rest at the initiation
of Roche-lobe overflow, shows that, indeed, at the initiation of mass
transfer, a $5 M_\odot$ primary (with a $3.84 M_\odot$ secondary) will begin
mass transfer with direct impact accretion on to the secondary star.

The second mechanism is associated with the phase-dependent Roche-lobe
overflow in an eccentric orbit binary.  We proposed in \S\ref{sec:parameters}
that HD~5501 is a binary which overfills its Roche lobe near periastron
passage, and underfills the Roche lobe near apastron.  This results in a
periodic variation in the mass transfer rate, otherwise known as
``phase-dependent Roche-lobe overflow.''   \citet{vanwinckel95} proposed
that such a situation could increase the eccentricity of the binary orbit
and \citet{soker2000} demonstrated theoretically the validity of that effect
for the case of an AGB star in a binary system that fills its Roche lobe
during the periastron passage (called the ``Soker Mechanism'').  

Thus the first mechanism (direct impact accretion) may be able to explain
how the
binary system could evolve from one with an initially circular orbit ($e = 0$)
to one with an elliptical orbit ($e > 0$).  The Soker mechanism
(phase-dependent Roche-lobe overflow) could then maintain or further increase
the eccentricity of the orbit.  However, this assumes that Roche-lobe
overflow begins while the orbit is circular.  The locus of points for that
lies precisely midway between the locus of points for periastron and apastron
Roche
lobe overflow (see Figure \ref{fig:HRD}).  Note that the solution represented
by the ``star'' in Figure \ref{fig:HRD} has not yet reached the point on its
evolutionary track where Roche-lobe overflow in a circular orbit would have
begun.  Indeed, very little of the solution space discussed in
\S \ref{sec:parameters} meets that criterion.  This suggests that the orbit
was elliptical before Roche-lobe overflow began.  But it is premature to draw
that conclusion because consideration of $\pm 2\sigma$ limits on the
astrometry/photometry box (the shaded box in Figure \ref{fig:HRD}) {\it does}
include solutions in which the primary could have first filled its Roche lobe
when the orbit was still circular.

\subsection{{\sc mesa} Binary Calculations}
\label{sec:MESAbin}

Further insight into HD~5501, in particular the questions posed in \S
\ref{sec:orbit_evolution} and \S \ref{sec:eccentricity} may be obtained by
modeling the system using the {\sc mesa} Binary package \citep{Paxton2015} which
allows
the user to evolve both stars in a binary system including mass transfer
between the components as well as mass loss from the system.  Eccentric orbits
may also be accommodated, and thus phenomena such as phase-dependent Roche-lobe
overflow may be investigated.  Eccentricity-modifying mechanisms, such as
those discussed in \S\ref{sec:eccentricity} as well as tidal circularization
can be included in the calculation.  However, because the package does not
incorporate a full-fledged dynamical calculation, it is not possible to
capture the full complexity of the HD~5501 system.

The {\sc mesa} binary package offers four different formulations for the mass
transfer scheme.  The first uses the scheme outlined by \citet{Ritter88} which
takes into account the finite scale height of the Roche-lobe-filling component.
This formulation is designed to be used in ``nearly semi-detached'' binaries.
The second uses the Kolb-Ritter formulation \citep{Kolb90} in which the radius
of the donor star is not constrained by its critical Roche-lobe radius.  The
third and the fourth formulations (``roche-lobe'' and ``contact'') both set the
mass transfer rate such that the donor star remains within its Roche lobe.
The calculations discussed below all use the ``Kolb-Ritter'' scheme as there
is little doubt that near periastron the donor star in HD~5501 overfills its
Roche lobe.

The {\sc mesa} binary package also allows the user to turn on the ``Soker
eccentricity enhancement'' mechanism arising from phase-dependent Roche-lobe
overflow which was discussed in \S \ref{sec:eccentricity}.  In addition,
the efficiency of mass transfer can be adjusted in the calculation using
the formulation of \citet{Tauris2006} which employs four different parameters,
$\alpha$, the fraction of mass lost from the vicinity of the donor via a fast
wind; $\beta$, the same for the accreting star; $\delta$, the fraction of
mass lost from a circumbinary planar toroid, and $\gamma$, the radius of
the circumbinary toroid.  The mass transfer efficiency is given by
$1 - \alpha - \beta - \delta$.  We set $\delta = 0$ in all of the calculations
reported below, but experimented with different values of $\alpha$ and
$\beta$, even though the primary, an A-type giant, is not expected to have
a substantial wind.  However, as will be seen in \S \ref{sec:mhalpha}, there
is the possibility for a mass outflow from the system occurring in the
vicinity of the $L_3$ point. Note that when the mass transfer efficiency is 100\% the evolution is conservative, whereas for lower efficiencies the evolution is non-conservative, meaning that both angular momentum and energy are lost from the system.

The two questions we are interested in addressing are 1) can the models
explain the fact that for HD~5501 $\dot{P} < 0$ (see \S \ref{sec:orbit})
and 2) the origin of the observed orbital eccentricity ($e \sim 0.24$).

We computed a number of sets of models all with different starting conditions
in order to explore some of the solution space.  All models discussed below
began the calculations at the zero-age main sequence (ZAMS) with beginning
masses $M_1 = 5.0M_\odot$ and $M_2 = 3.84M_\odot$ and period = 7.531 days.
Different assumptions were made for the initial eccentricity of the orbit as
well as the efficiency of mass transfer.  The {\sc mesa} binary calculations
evolved both stars simultaneously and calculated the period and eccentricity
of the orbit, the masses of the stars, and the mass transfer rate at each
point in time.  The calculations were carried through the entire period of
rapid mass transfer via Roche-lobe overflow. 

\begin{figure} 
\includegraphics[width=3.2in,angle=0]{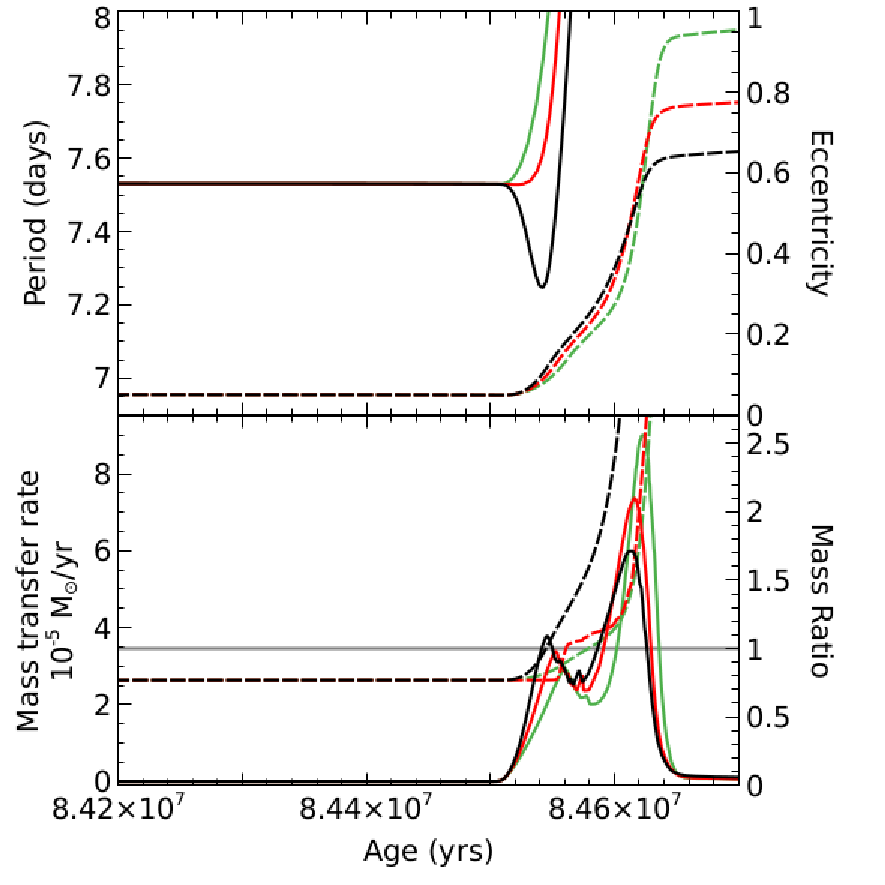} 
\caption{The results for a {\sc mesa} binary calculation for HD~5501 with an
  initial orbital eccentricity $e = 0.05$ and with the Soker mechanism
  turned on.  The upper panel
  shows the results for the evolution of the orbital period (solid lines) and
  the eccentricity (dashed lines) for mass transfer efficiencies of
  100\% (black), 50\% (red) and 1\% (green).  The lower panel shows the
  results for the mass transfer rate (solid lines) and the mass ratio (dashed
  lines).  The colour coding is the same as the top panel.  The gray horizontal
  line shows the location of the mass ratio = 1.}
\label{fig:e05}
\end{figure}

The first model begins with an initial small eccentricity,  $e = 0.05$.
With the Soker mechanism
turned off, the eccentricity remains essentially constant during the
integration.  However, with the Soker mechanism turned on
(see Fig \ref{fig:e05}), the eccentricity rises rapidly during the period
of rapid mass transfer.  An important feature of the calculation is that the
period derivative turns negative right at the beginning of rapid mass
transfer via Roche-lobe overflow and remains negative for about 30,000 years.
This occurs whether or not the Soker mechanism is turned on, although with
the Soker mechanism turned off $\dot{P}$ remains negative for slightly
longer $\sim$ 35,000 years.  This period of negative $\dot{P}$ is relevant
to HD~5501 as we have observationally demonstrated
that $\dot{P} < 0$, and this strongly suggests that we are observing
HD~5501 {\it shortly after the onset of mass transfer via Roche-lobe overflow}.
Indeed, the model $\dot{P} \sim -4.2 \times 10^{-8}$ days/day is reasonably
close to the observed $\dot{P} = -1.2 \times 10^{-7}$ (see \S \ref{sec:orbit}). 
The interval of time during which $\dot{P} < 0$ coincides
with the length of time that it takes for the mass ratio to evolve from
its original value (0.768) to unity.  Once that occurs, $\dot{P}$ turns
sharply positive, and by the end of Roche-lobe overflow, the orbital period has
increased to about 650 days.  In addition, during the period of rapid mass
transfer, the eccentricity rises to about 0.65.

Can this rapid rise in the eccentricity explain the current eccentricity of
the HD~5501 system?  According to the calculation, during the interval when
$\dot{P} < 0$, the eccentricity rises only to 0.12, and that eccentricity is
only achieved when the mass ratio passes through unity.  Most of the increase
in the eccentricity occurs when the donor star is the least massive, which
does not appear to be the case for HD~5501.  The current observed mass ratio
of HD~5501 ($q \sim 0.77$) in this scenario implies, once again, that HD~5501
has just started rapid Roche-lobe overflow.  This model thus implies that the Soker mechanism, at least as
implemented in the {\sc mesa} binary code, is incapable of explaining the current
observed eccentricity ($e \sim 0.24$).  Indeed, to achieve $e = 0.24$ by the
time that $q = 1$, the initial eccentricity must be $\sim 0.17$, which is nearly
as difficult to explain as an initial eccentricity of 0.24.  This suggests
that the Soker mechanism implementation in the {\sc mesa} binary code
requires revision, or that there is some other physical mechanism, such as
a third body in the system, that explains the high eccentricity or that the actual eccentricity of the orbit is more moderate.  Another
possibility is that
the binary was formed with a high eccentricity orbit; the models indicate that
the eccentricity-modifying mechanisms are ineffective at modifying the initial
eccentricity until mass transfer begins.

Changing the initial eccentricity in the model does not materially alter
the results reported above.  However, changing the assumptions about the
efficiency of mass transfer does affect
the results.  In particular, reducing the mass transfer efficiency both
reduces the length of time that $\dot{P}$ remains negative and makes that
derivative more shallow.  Once a mass transfer efficiency of 50\% is reached,
that interval at the start of mass transfer during which $\dot{P} < 0$ is
eliminated.  However, these models with lower mass transfer efficiency
end up with very high final eccentricities.  At 50\% efficiency, the final
eccentricity of the binary orbit is nearly 0.8.  Reducing
the efficiency further to 1\% ends with a near disruption of the binary
($e \sim 1$),
presumably because almost all the mass is lost from the system via a massive
wind.  Those scenarios are probably physically unreasonable for the
HD~5501 system.

\subsection{A first step toward modeling the complex behavior of the
  H~$\alpha$ line}
\label{sec:mhalpha}

The complex variation observed at H~$\alpha$ is described and illustrated in
\S \ref{sec:halpha}. That line profile shows a predominant emission peak that
persists at all phases but also shows a roughly sinusoidal variation in
velocity (see Figures \ref{fig:halpha} and \ref{fig:Havel}) that does not
coincide with the radial velocity solution derived in \S \ref{sec:RV} based on
photospheric lines.  Near phases 0.6 -- 0.8
the H~$\alpha$ emission can show a double peak with a variable V/R ratio.
In addition, the profile can show, especially near secondary eclipse,
highly variable absorption components, some with blue-shifted velocities as
high as 500 km s$^{-1}$ relative to the systemic velocity.

In this section we attempt to model the predominant H~$\alpha$ emission peak.
The understanding gained through attempts at modeling the system (see
\S \ref{sec:lcurve} and \S \ref{sec:MESAbin}) that the primary star
significantly overfills its Roche lobe near periastron passage suggests
very strongly the possibility of significant mass loss from the system
through either or both Lagrange points $L_2$ and $L_3$.  Our hypothesis is
that the predominant emission peak results from outflow from the primary
in the vicinity of the $L_3$ point.

We begin our analysis with the {\sc shellspec} light curve model of
\S \ref{sec:lcurve}.  The {\sc shellspec} code enables two ways of modeling localized outflows
from a binary
system.  The outflow may be defined using a {\it jet} ``object'' or a
{\it stream} ``object''.  The main difference between the two (at least for the
purposes of this model) is that the
{\it jet} object allows diverging stream lines in the outflow whereas in the
{\it stream} object those stream lines are parallel, even though the
cross-sectional radii
at the beginning and end of the stream may be different.  In both cases, the
gas density, which is specified at the beginning of the flow, is scaled
along the flow to satisfy the continuity equation.  We actually end up with
quite similar results independent of which object we choose to employ.
{\sc shellspec}
performs a radiative transfer calculation for the objects added to the
system, so it is possible to calculate the contribution to the H~$\alpha$
profile from those objects.  We have modelled outflows in the vicinity of
the $L_2$ and $L_3$ points in the following way:

The outflow from the vicinity of the $L_2$ point was modelled with a
{\it stream} object originating at the $L_2$ point and oriented in the
average direction of particle stream lines escaping from the system at rest
at $L_2$.

\begin{figure}
  \includegraphics[width=2.0in]{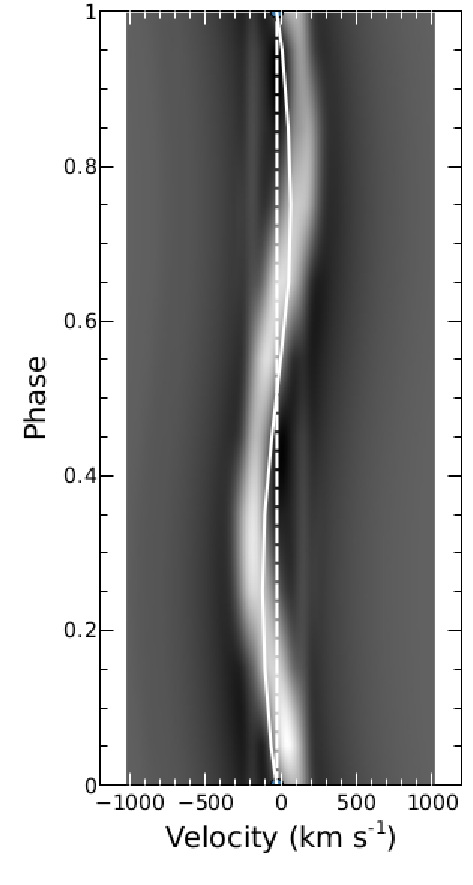}
  \caption{A grayscale rendition of synthetic spectra at H~$\alpha$ as a function
    of photometric phase produced by
    {\sc shellspec} for a model of the HD~5501 system including outflow
    from the primary star in the vicinity of the Lagrange $L_3$ point.  Compare
    with Figure \ref{fig:halpha}.  The solid white line is the radial velocity
    curve of the primary {\it determined from the synthetic spectra} produced
    by {\sc shellspec} via cross correlation.  The dashed white line is the
    systemic velocity measured from those synthetic spectra. These are included
    to show that the model has the same orbital direction as the real star
    (again, compare with Figure \ref{fig:halpha}a).
    The model also contains a minor
    contribution from a circumbinary disc which shows up as faint constant
    velocity emission at about +150 and -200 km s$^{-1}$.  See text for
    more details.  This figure is based on continuum-normalized synthetic
  spectra.  It is scaled similarly to Figure \ref{fig:halpha}.}
  \label{fig:L3of}
\end{figure}

Since the outflow from the vicinity of the $L_3$ point presumably
originates near the surface of the primary during periastron passage, that
outflow was modelled as a truncated one-sided conical jet with apex at the
centre of the primary and oriented in the average direction of particle
streamlines originating from points near the surface of the primary and
escaping from the system in the vicinity of $L_3$.  For both the
$L_2$ and $L_3$ objects, outflow temperatures, densities, velocities and
opening angles were varied in an attempt to reproduce the observations
within the limitations of the model.  We used the following constraints on
the outflow parameters:
$T_{\rm outflow} < T_{\rm eff}({\rm primary})$ and
$\rho_{\rm outflow} \ll \rho_{\rm photosphere}({\rm primary})$.  The outflow
velocities
were guided by the particle streamline velocities.  Both stream and jet
were truncated near the edge of the computation grid (at $70 R_\odot$ from the
point of origin).

It was found that the predominant emission peak can only be explained by
an outflow from the vicinity of the $L_3$ point (see Figure \ref{fig:L3of})
which is in good qualitative agreement with the observed emission peak (see
Figure \ref{fig:halpha}).  The outflow from $L_2$ produces an emission peak that has a radial velocity
variation that is $180^\circ$ out of phase with the $L_3$ outflow and thus
is inconsistent with the observations.  That does not mean, however, that
there is no outflow from $L_2$.  Indeed, the violet peak of the double-peaked
phase of H~$\alpha$ may derive from an outflow from the vicinity of $L_2$.
However, if that is the case, we should also see double-peaked emission at the
symmetrical phase in the orbit: phases 0.2 - 0.4, which we do not.
Emission from the inner
hotter part of a circumbinary shell or disc can also lead to double-peaked
emission.  Indeed, Figure \ref{fig:L3of}
includes the effects of a low density circumbinary disc which appear as a
double-peaked structure at the required phases.  That
    circumbinary disc is centred on the centre of mass of the system and rotates
    with Keplerian velocities.  It has an outer radius of $45 R_\odot$ (and thus models only the hot inner part of the disc due to limitations of the computational grid), a uniform gas density of $3.0 \times 10^{-15}$~g~cm$^{-3}$ and a
uniform temperature of 8000K.  The effects of that disc can be seen in Figure \ref{fig:L3of} as two faint emission peaks at constant and equal red and blueshifts relative
to the systemic velocity.  As a result, that circumbinary
disc also displays double-peaked emission at phases 0.2 - 0.4.  The fact
that we do
not see double-peaked emission at those phases may have to do with the
phase-dependent nature of the mass transfer from the primary to the
secondary or with a non-homogeneity in the circumbinary disc
(see Figure \ref{fig:EQW}).
The fact that the V/R ratio of that double-peaked emission varies
may be due to a temporary enhancement in the density of
the $L_2$ outflow or changes in the density of the circumbinary disc.
Investigating those details will be the subject for further study.

According to numerical calculations, the mass outflow will, in a coordinate
system rotating with the binary, spiral outwards from the system and
presumably form or become incorporated into the circumbinary disc.  The
observed high-velocity
H~$\alpha$ absorption components likely derive from this extended outflow when
seen in absorption against the primary star.  Further analysis of
this system with high-resolution spectra should help to elucidate this
phenomenon as well as the many questions remaining about this astrophysically
interesting system.

While this model explains the predominant emission feature
  of the
  H~$\alpha$ profile qualitatively, it is important to investigate whether the
  outflow from the vicinity of $L_3$ materially affects the Johnson $V$-band
  light curve
  modelled in \S \ref{sec:lcurve}.  To investigate this, we have recomputed
  the model light curve for the system including the outflow from the
  vicinity of $L_3$ and the circumbinary disk.  The resulting light curve is
  almost identical to the modelled light curve in \S \ref{sec:lcurve}.  This
  suggests that the outflows and the circumbinary disk have a negligible
  effect on the $V$-band light curve, one of the assumptions that we made in
  \S \ref{sec:lcurve}.

\section{Conclusions}

The following are the main conclusions of this paper:

\begin{enumerate}
\item HD~5501 is an interacting eclipsing binary star with a B9/A0 III primary and a
  presumed B7 V secondary.  No sign of the secondary is apparent in our spectra.
  The spectrum of the primary exhibits characteristics of a shell star as well as strong and variable emission and absorption at H~$\alpha$.
\item The location of the primary star in the HR Diagram suggest that it has
  undergone hydrogen core exhaustion and is evolving rapidly across the Hertzsprung Gap.
\item The orbital period ($\sim 7.5$ days) is decreasing with the remarkably
  rapid time-scale
  for orbital evolution $P/\dot{P} \approx 170,000$ years or less.
\item The light curve of HD~5501 is peculiar in the sense that the shapes,
  depths and even timings of the primary and secondary eclipses change from
  orbit-to-orbit.  These changes appear to be caused by dynamical chaos
  perhaps associated with a variable shape and size of an accretion torus
  around the secondary.
\item The presence of shell lines in the spectrum of the primary, reddening
  greater than line-of-sight values, as well as circumbinary
  absorption components in the cores of strong lines such as Ca~{\sc ii} K suggest
  the presence of a circumbinary shell or disc.
\item The orbit of the binary is somewhat eccentric with an eccentricity possibly as large as $0.24$. Our
  analysis implies that the primary star overfills its Roche lobe near
  periastron.  Modeling with the {\sc mesa} Binary code suggests that the system
  has only recently begun Roche-lobe overflow.
\item Analysis of the highly variable H~$\alpha$ profile indicates that the
  system is losing mass through the Lagrange $L_3$ point.
\item HD~5501 is an astrophysically interesting system which may yield valuable
  information about binary star evolution at the onset of Roche-lobe overflow,
  as well as insights into eccentricity-modifying mechanisms such as the
  Soker mechanism.
\end{enumerate}
\section*{Acknowledgements}

We thank the referee, Dr. Douglas Gies, whose comments led to significant improvements to this paper.

This work has made use of data from the European Space Agency (ESA) mission
{\it Gaia} (\url{https://www.cosmos.esa.int/gaia}), processed by the {\it Gaia}
Data Processing and Analysis Consortium (DPAC,
\url{https://www.cosmos.esa.int/web/gaia/dpac/consortium}). Funding for the DPAC
has been provided by national institutions, in particular the institutions
participating in the {\it Gaia} Multilateral Agreement.

This work has used observations acquired at the Vatican Observatory Advanced Technology Telescope, Mt. Graham, Arizona.

This research has made use of the NASA Exoplanet Archive, which is operated by the California Institute of Technology, under contract with the National Aeronautics and Space Administration under the Exoplanet Exploration Program.

We acknowledge with thanks the variable star observations from the AAVSO International Database contributed by observers worldwide and used in this research.

The Harvard plates were measured as part of the just-completed Digital
Access to a Sky Century @ Harvard (DASCH, J. Grindlay PI), available at
\url{https://dasch.cfa.harvard.edu/}, which has been partially supported
by NSF grants AST-0407380, AST-0909073, and AST-1313370.

This work made use of the ``Modules  for Experiments in Stellar Astrophysics''
\citep[{\sc mesa}:][]{Paxton2011, Paxton2013, Paxton2015, Paxton2018, Paxton2019},
  specifically version 24.08.1, and version 24.7.1 of the {\sc mesasdk}.

This work made use of {\sc astropy} (\url{http://www.astropy.org}): a community-developed core Python package and an ecosystem of tools and resources for astronomy \citep{astropy2013, astropy2018, astropy2022}. 

This research made use of {\sc ccdproc}, an Astropy package for
image reduction \citep{matt_craig_2017_1069648}.

This research made use of {\sc photutils}, an Astropy package for
detection and photometry of astronomical sources
\citep{larry_bradley_2023_7946442}.

\section*{Data Availability}

Photometric and spectroscopic data obtained at the Dark Sky Observatory
will be made available upon reasonable request to the corresponding
author, Richard Gray (grayro@appstate.edu).  Likewise, spectroscopic
data obtained with the Vatican Advanced Technology Telescope will be made
available upon reasonable request to Christopher Corbally (cjc@arizona.edu).
Queries for use of the Harvard Plate archive HD~5501 data should be
addressed to Bradley Schaefer (bradschaefer@me.com); the DASCH data may be
publicly accessed at \url{http://dasch.rc.fas.harvard.edu/lightcurve.php}.
AAVSO photometric data obtained for this paper are available in the AAVSO
International Database
(\url{https://www/aavso.org}).  Spectroscopic data obtained by AAVSO
members will also be available through that site.  BAA (British Astronomical
Association) photometric data obtained for this paper are available in the
BAA Photometry Database (\url{https://britastro.org/photdb}).
Spectroscopic data obtained by BAA members will also be available in the
BAA Spectroscopy Database (\url{https://britastro.org/specdb}).



\bibliographystyle{mnras}
\bibliography{gray_mnras} 





\bsp	
\label{lastpage}
\end{document}